\documentclass[sigconf]{acmart}

\usepackage{multirow}
\usepackage{makecell} 
\usepackage{xcolor}
\usepackage{booktabs}
\usepackage{colortbl}
\usepackage{listings}

\usepackage{amsmath}
\usepackage{amsfonts}
\usepackage{booktabs}
\usepackage{adjustbox}
\usepackage{epstopdf}
\usepackage{subcaption}
\usepackage{tabulary}
\usepackage[inline]{enumitem}
\AtBeginDocument{%
  }

\copyrightyear{2026}
\acmYear{2026}
\setcopyright{cc}
\setcctype{by}
\acmConference[SIGIR '26] {Proceedings of the 49th International ACM SIGIR Conference on Research and Development in Information Retrieval}{July 20--24, 2026}{Melbourne, VIC, Australia.}
\acmBooktitle{Proceedings of the 49th International ACM SIGIR Conference on Research and Development in Information Retrieval (SIGIR '26), July 20--24, 2026, Melbourne, VIC, Australia}
\acmISBN{979-8-4007-2599-9/2026/07}
\acmDOI{10.1145/3805712.3809532}
\begin{document}

\title{Attention Grounded Enhancement for Visual Document Retrieval}

\author{Wanqing Cui}
\email{cuiwanqing.cwq@taobao.com}
\orcid{0000-0001-5015-5252}
\affiliation{%
  \institution{Alibaba Group}
  \city{Beijing}
  \country{China}
}

\author{Wei Huang}
\email{huangwei21@mails.ucas.ac.cn}
\affiliation{%
  \institution{University of Chinese Academy of Sciences}
  \city{Beijing}
  \country{China}
}

\author{Yazhi Guo}
\email{guoyazhi.gyz@taobao.com}
\affiliation{%
  \institution{Alibaba Group}
  \city{Beijing}
  \country{China}
}

\author{Yibo Hu}
\email{boxuan.hyb@taobao.com}
\affiliation{%
  \institution{Alibaba Group}
  \city{Beijing}
  \country{China}
}

\author{Meiguang Jin}
\email{meiguang.jmg@taobao.com}
\affiliation{%
  \institution{Alibaba Group}
  \city{Beijing}
  \country{China}
}

\author{Junfeng Ma}
\email{jack.majf@taobao.com}
\affiliation{%
  \institution{Alibaba Group}
  \city{Hangzhou}
  \country{China}
}

\author{Keping Bi}
\authornote{Corresponding author.}
\email{kepingbi@acm.org}
\affiliation{%
  \institution{University of Chinese Academy of Sciences}
  \city{Beijing}
  \country{China}
}


\begin{abstract}

Visual document retrieval requires understanding heterogeneous and multi-modal content to satisfy implicit information needs. Recent advances use screenshot-based document encoding with fine-grained late interaction to encode holistic information and capture nuanced alignments, significantly improving retrieval performance. However, retrievers are still trained with coarse global relevance labels, without revealing which regions support the match. As a result, retrievers tend to rely on surface-level cues and struggle to capture implicit semantic connections, hindering their ability to handle non-extractive queries.
To improve fine-grained relevance modeling,  we propose a \textbf{A}ttention-\textbf{G}rounded \textbf{RE}triever \textbf{E}nhancement (AGREE) framework. AGREE leverages cross-modal attention from multimodal large language models (MLLMs) as proxy supervision to guide the retriever in identifying relevant document regions. Specifically, AGREE extracts attention maps from the MLLM that highlight which document regions are attended to based on the query. These attention scores serve as local, region-level relevance signals. During training, AGREE combines local signals with the global document-level relevance label to jointly optimize the retriever. This dual-level supervision enables the model to learn not only whether documents match, but also which content drives relevance. 
Experiments on the challenging visual document retrieval benchmark, ViDoRe V2, show that AGREE significantly outperforms the global-supervision-only baseline by 12.82\% and 5.03\% in terms of average nDCG@1 and nDCG@5. 
Quantitative and qualitative analyses further demonstrate that AGREE promotes deeper alignment between query terms and document regions, moving beyond surface-level matching toward more accurate and interpretable retrieval. Our code is available at: \url{https://github.com/VickiCui/AGREE}.

\end{abstract}







\maketitle

% related work
% 1. visual document （i.e. document screenshots） retrieval
% 2. matching with late interaction
% 3. attention distillation

% method
% 对于相关文档，模型看哪是明确的。对于不相关的文档，很难给出明确的指导。因此只加正例的attention。

% experiment
% 除了主实验、另外分析了：1. 不同attention gather方法的效果；2. 高注意力比例的影响、loss权重；3. 引入不同注意力损失样本的比例的影响、以及优先引入难样本注意力损失的影响；4. 不同难易度样本attention差异 5. case study

\section{Introduction}

Visual document retrieval~\cite{faysse2024colpali, macé2025vidorebenchmarkv2raising, nomicembedmultimodal2025} aims to find document pages that are semantically relevant to a given query. These pages often combine text, figures, tables, and layout structures, forming rich, heterogeneous content. Visual document retrieval is a critical component in real-world applications such as retrieval-augmented generation (RAG)~\cite{lewis2020retrieval}. In scenarios like financial reports, scientific papers, and technical manuals, key evidence is frequently distributed across visual and textual elements. Effective visual document retrieval enables models to ground in these complex sources, ensuring accurate and context-aware information response.

To preserve holistic information, recent visual document retrieval methods encode document pages as screenshots and follow a text-to-visual retrieval paradigm~\cite {faysse2024colpali, ma2024unifying}. Such approaches outperform text-centric methods~\cite{luo2023unifying, liu2023mmhqa, yu2023unified}, which use OCR and image captioning to extract text before applying standard text retrieval models. Notably, screenshot-based visual document retrieval is significantly more challenging than traditional image retrieval. Document screenshots contain high information density, with tightly coupled multimodal content that requires joint interpretation, requiring fine-grained alignment. Moreover, queries in visual document retrieval often arise from complex information needs, requiring semantic reasoning to identify implicitly relevant content beyond keyword matching.

\begin{figure}[t]
  \centerline{\includegraphics[width=1.0\linewidth]{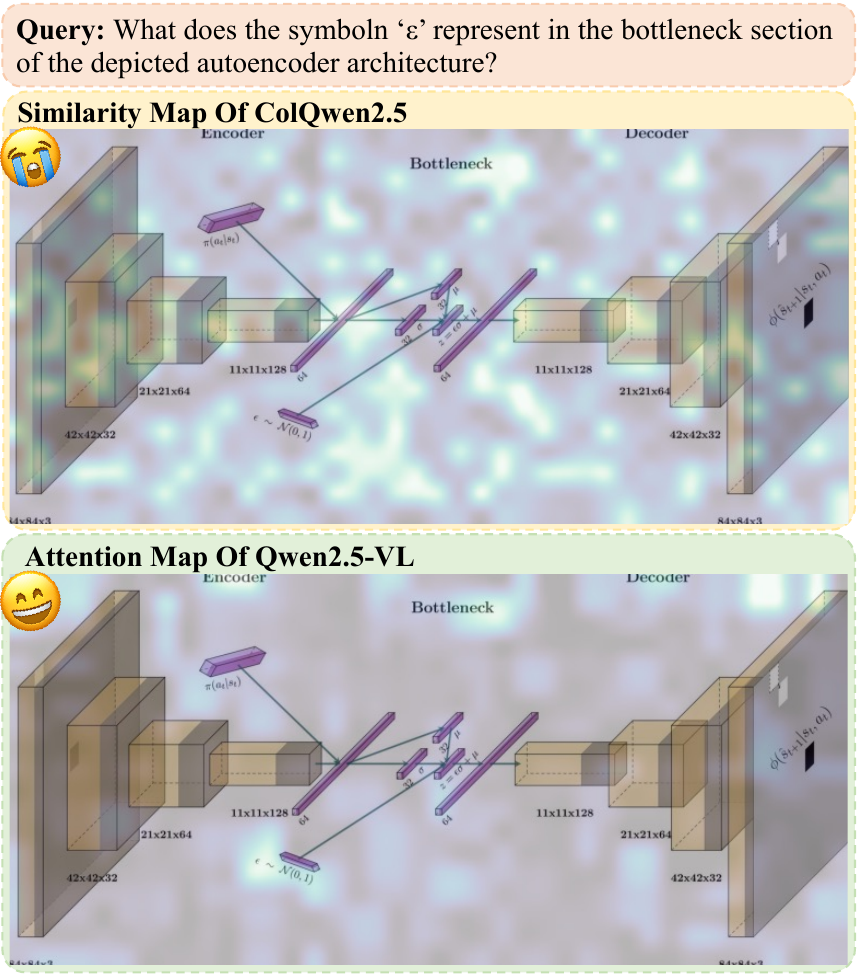}}
  \caption{The similarity map of ColQwen2.5 (top),  versus the query-to-image attention map from Qwen2.5-VL (bottom).}
  \label{fig:intro}
  % \vskip -0.2in
\end{figure}

% \begin{figure*}[t]
%   \centerline{\includegraphics[width=0.8\linewidth]{figures/intro.pdf}}
%   \caption{The similarity map of ColQwen2.5 (left),  versus the query-to-image attention map from Qwen2.5-VL (right).}
%   \label{fig:intro}
% \end{figure*}

Despite screenshot-based methods can encode more comprehensive information, they typically rely on global embeddings that compress all content into a single vector, struggle to model fine-grained alignments. To overcome this limitation, state-of-the-art visual document retrieval method, i.e., ColQwen~\cite{faysse2024colpali}, adopts fine-grained late interaction~\cite{khattab2020colbert}. It preserves token-level and patch-level embeddings and computes relevance using the MaxSim operator: matching each query token to its most similar image patch and aggregating the maximum similarities. This enables detailed alignment between specific query terms and relevant regions in the document page.

However, current late interaction retrievers are trained using only global binary relevance labels. These labels only indicate whether a page is relevant, but provide no insight into which regions support the match. Therefore, when the matching clues are implicit, it is difficult for the retriever to learn why a page is relevant under such coarse supervision. As a result, the retriever still struggles to tackle non-extractive queries. As shown in Figure~\ref{fig:intro}, the query matches the page not only on explicit terms like ``bottleneck'' and ``$\epsilon$'', but also through an implicit link from ``autoencoder'' to ``encoder-decoder''. However, under global supervision, ColQwen’s similarity map highlights only ``bottleneck'', overlooks ``$\epsilon$'' due to its small font size, not to mention the implicit correspondence. This highlights the need for fine-grained supervision that can directly indicate the matching clues. While human annotations could provide such signals, it is labor-intensive and impractical to scale. Therefore, scalable sources of proxy supervision are essential.

Multimodal large language models (MLLMs)~\cite{bai2025qwen25vltechnicalreport, liu2023llava, liu2024llavanext} exhibit precisely the kind of fine-grained alignment we seek. Recent studies~\cite{zhang2025mllms, wang2025exploring} have demonstrated that attention patterns of MLLMs consistently highlight semantically critical regions. As shown in Figure~\ref{fig:intro}, based on the query, an MLLM attends not only to exact keyword matches, but also to implicit associations. Such attention weights therefore has the potential to provide fine-grained signals, without costly manual annotation.

Based on these observations, we propose the \textbf{A}ttention \textbf{G}rounded \textbf{RE}triever \textbf{E}nhancement (AGREE) training framework to improve performance on visual document retrieval tasks that require implicit, non-extractive matching. AGREE leverages fine-grained supervision through three key stages:
(1) MLLM Attention Annotation: We employ a pretrained MLLM to generate query-conditioned, patch-level attention maps.
% (2) Spatial-Preserving Attention Downsampling: To prevent blurry images from hindering accurate character recognition and ensure the quality of attention, we use more patches during attention extraction than relevance matching. We then downsample the attention maps via area-based pooling to align with the retriever’s patch layout.
(2) Spatial-Preserving Attention Downsampling: To reconcile the high resolution required for MLLM reasoning with the lower resolution used for efficient retrieval, we extract attention from high-resolution inputs and downsample the resulting maps to match the retriever's patch grid.
(3) Attention-Guided Training: The retriever is trained with a dual objective: (i) a global contrastive loss for page-level relevance, (ii) and a local alignment loss that encourages the retriever's patch similarity scores to mimic the MLLM attention on positive pairs. This dual-level supervision enables the model to learn not only whether documents match, but also which content drives relevance. 
Crucially, by aligning with the attention of a large MLLM, AGREE distills the teacher's intrinsic reasoning capabilities into the retriever. This facilitates the model to generalize beyond the simple extractive patterns of the training data and handle the complex, implicit queries. 
% found in benchmarks like ViDoRe V2 \cite{macé2025vidorebenchmarkv2raising}.

% Evaluation demonstrates that AGREE significantly improves retrieval performance on questions beyond non-extractive matching across various backbones. On ViDoRe V2 benchmark~\cite{macé2025vidorebenchmarkv2raising}, where relevance requires semantic reasoning beyond keyword matching, AGREE achieves substantial improvements: average nDCG@1 increases from 0.5481 to 0.6184 (+12.82\%), and average nDCG@5 from 0.5859 to 0.6154 (+5.03\%), demonstrating its ability to capture implicit, non-extractive alignments. On the easier visual document retrieval benchmark, ViDoRe V1~\cite{faysse2024colpali}, where retrieval is dominated by explicit text overlap, AGREE maintains competitive performance, confirming its robustness across varying levels of retrieval complexity. Further quantitative and qualitative analysis reveals that the retriever trained with AGREE attends to a broader set of relevant regions and captures implicit correspondences. This indicates that AGREE enables the retriever to move beyond surface-level matching toward more interpretable, region-aware retrieval. 
Evaluation shows that AGREE consistently improves retrieval performance on questions requiring non-extractive semantic matching, across multiple backbones. On ViDoRe V2~\cite{macé2025vidorebenchmarkv2raising}, where relevance depends on semantic reasoning beyond keyword overlap, AGREE yields substantial gains: average nDCG@1 improves from 0.5481 to 0.6184 (+12.82\%), and average nDCG@5 increases from 0.5859 to 0.6154 (+5.03\%), highlighting its ability to capture implicit, non-extractive alignments. On the easier ViDoRe V1 benchmark~\cite{faysse2024colpali}, where retrieval is largely driven by explicit textual overlap, AGREE remains competitive, confirming its robustness across varying levels of retrieval difficulty. Further quantitative and qualitative analyses indicate that AGREE-trained retrievers attend to a broader set of relevant regions and better capture implicit correspondences, enabling retrieval to move beyond surface-level matching toward more interpretable, region-aware behavior.
% Moreover, the effectiveness of AGREE is closely tied to the quality of the fine-grained labels, suggesting that improvements in grounding signals will further boost performance.

Overall, our main contributions are summarized as follows:
% \begin{itemize}
% \item We propose AGREE, a novel training framework that uses MLLM attention as fine-grained supervision to improve visual document retrieval performance on implicit matching.
% \item Extensive experiments across multiple backbones demonstrate that AGREE achieves significant gains on challenging, non-extractive retrieval tasks, demonstrating its effectiveness and adaptability.
% \item Quantitative and qualitative analyses show that AGREE shifts retrieval from keyword matching to rationale-aware alignment, fundamentally reshaping how retrievers understand relevance.
% \end{itemize}
% (1) We propose AGREE, a novel training framework that uses MLLM attention as fine-grained supervision to improve visual document retrieval performance on implicit matching.
(1) We propose AGREE, a novel training framework that distills query-conditioned MLLM attention into late-interaction visual document retrievers, providing scalable fine-grained supervision for implicit relevance matching.
(2) Extensive experiments across multiple backbones demonstrate that AGREE achieves significant gains on challenging, non-extractive retrieval tasks, demonstrating its effectiveness and adaptability.
% (3) Quantitative and qualitative analyses show that AGREE shifts retrieval from keyword matching to rationale-aware alignment, fundamentally reshaping how retrievers understand relevance.
(3) Quantitative and qualitative analyses show that AGREE encourages retrievers to move beyond keyword overlap by attending to a broader set of relevant regions and capturing implicit query–document correspondences.
\section{Related Work}

\subsection{Visual Document Retrieval}
Traditional document retrieval systems primarily rely on textual content, a characteristic shared by both bag-of-words methods~\cite{sparck1972statistical, robertson2009probabilistic} and modern neural retrieval approaches~\cite{reimers2019sentence, karpukhin2020dense, wang2022text}. However, real-world documents are inherently multimodal, often combining text with images, tables, and other visual elements. Consequently, recent research has focused on incorporating such non-textual information to enhance retrieval. A common strategy involves separately extracting and encoding visual and textual features from documents~\cite{yasunaga2022retrieval, chen2024vtqa, wei2024uniir, jiang2024vlm2vec, tanaka2023slidevqa}. 

A key limitation of these methods is their potential difficulty in capturing complex cross-modal interactions when modalities are processed in isolation. An alternative approach~\cite{luo2023unifying, liu2023mmhqa, yu2023unified} focuses on generating a unified textual representation: the textual content is extracted using Optical Character Recognition (OCR)~\cite{memon2020handwritten, smith2007overview} or PDF parsing~\cite{pymupdf, pdfminer}, and specialized models generate descriptive captions for the non-textual elements. However, converting rich non-textual information into textual descriptions inherently risks significant information loss. To address these challenges, concurrent works~\cite{faysse2024colpali, ma2024unifying, cho2024m3docrag} directly encode the screenshot of a document for retrieval, aiming to preserve the holistic multimodal structure and inherent relationships between elements.

\subsection{Fine-grained Image Retrieval}

Fine-grained matching, or late-interaction retrieval, has emerged as a prominent paradigm in information retrieval, offering an alternative to single-vector representations by capturing fine-grained term-level interactions. This approach, pioneered by ColBERT~\cite{khattab2020colbert}, computes relevance by decomposing queries and documents into token-level embeddings and performing an aggregation operation with MaxSim. Its effectiveness has led to its adoption in multimodal contexts, such as image-text matching~\cite{yao2021filip, asokan2025finelip, bica2024improving} and visual document retrieval~\cite{faysse2024colpali}. 
However, encoder models like CLIP~\cite{radford2021learning} and SigLIP~\cite{zhai2023sigmoid} are typically pre-trained using only global embedding alignment, without explicit optimization of token-level embedding. As a result, their token embeddings may lack the discriminative power required for. To enhance the quality of local embeddings, subsequent approaches have augmented global contrastive learning with auxiliary supervision that aligns regional text descriptions with corresponding image regions~\cite{xie2025fg, jing2024fineclip}.

% This "detect-then-describe" strategy is suboptimal for VDR. Unlike in natural images, salient information in visual documents is often embedded in text blocks, tables, or figures. These regions are not conventional "objects" and are difficult to reliably detect or describe with discrete text. To address this challenge, we propose a novel knowledge distillation framework. Instead of generating explicit labels, our method distills supervision from a teacher MLLM in the form of continuous attention maps. This provides a soft region-level supervisory signal that directly guides the optimization of the late-interaction mechanism, bypassing the brittle process of region proposal and captioning. Our approach is therefore a more direct and robust solution tailored to the specific challenges of visual document retrieval.

\subsection{Knowledge Distillation}
Knowledge distillation~\cite{hinton2015distilling} aims to transfer knowledge from a powerful teacher model to a compact student model by encouraging the student to mimic the teacher’s output distributions. In information retrieval, KD has been widely adopted to distill large, complex retrievers into smaller, efficient counterparts~\cite{rao2023dynamic, csizmadia2025distill}, while also enabling the student to learn nuanced soft labels that encode richer information than binary hard labels~\cite{lu2020twinbert, lin2021batch, dong2023dual, ma2023using, miech2021thinking, izacard2020distilling, zhou2024fine, li2024intermediate}.

Attention-based knowledge distillation~\cite{zagoruyko2016paying, sau2021deep, shin2022teaching, jin2024align} extends knowledge distillation to intermediate representations, which encourages the student to mimic not only the teacher's prediction results but also the teacher's internal attention patterns. These methods typically align attention maps across layers to regularize the student’s reasoning process.

% While we also exploit attention as a source of supervision, our objective diverges fundamentally: rather than guiding the student’s internal computation, we treat the attention as a mechanism for \emph{generating fine-grained alignment labels}, i.e., what specific content matches the query. 
Difference from AGREE. Unlike traditional KD, which focuses on model compression by mimicking internal feature activations, AGREE addresses a weak supervision challenge. We treat the MLLM not as a teacher to be compressed, but as an annotator that synthesizes missing \emph{fine-grained labels}. Thus, AGREE aligns the retriever’s similarity scores with MLLM attention, rather than aligning intermediate feature maps.
\begin{figure*}[t]
  \centerline{\includegraphics[width=1.\linewidth]{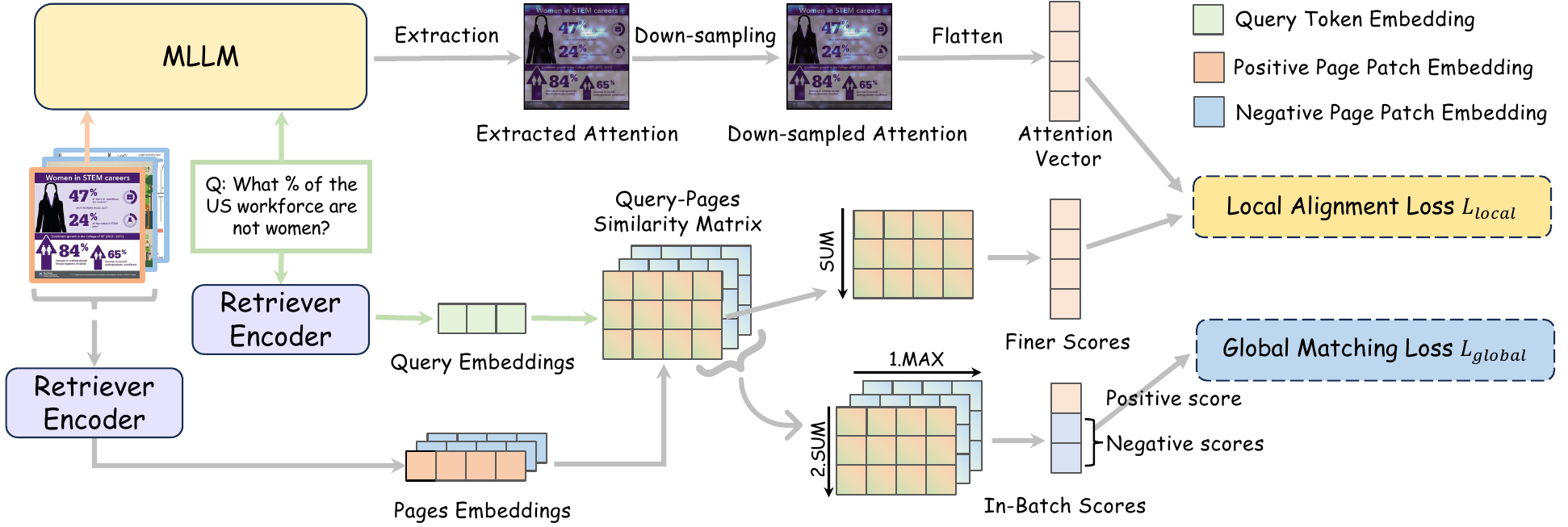}}
  \caption{
  % Overall pipeline of the proposed AGREE framework: A positive query–document pair is fed into a frozen MLLM, and the model first extracts high-resolution patch-level attention and then downsamples it to a attention vector. In parallel, the retriever encoder produces query and document patch embeddings, from which we obtain (i) a global relevance score for in-batch contrastive learning and (ii) a patch-level similarity vector, \emph{only for positive pairs}.  The latter is aligned with the MLLM attention through an attention-guided loss, while the former is optimized with a contrastive loss.
  Overview of the AGREE training framework. During training, a frozen MLLM processes only positive query–page pairs to generate high-resolution, query-conditioned attention maps over the page screenshot. These maps are downsampled via area-based pooling to align with the retriever's patch layout. In parallel, the retriever encoder computes embeddings for both query and candidate pages, from which we obtain (i) a global relevance score for in-batch contrastive learning and (ii) a patch-level similarity vector, \emph{only for positive pairs}. The latter is aligned with the MLLM attention through an attention-guided loss, while the former is optimized with a contrastive loss.
  }
  \label{fig:method}
  % \vskip -0.15in
\end{figure*}

\section{Preliminaries}
\label{sec:preliminaries}
\paragraph{Problem Formulation.}
The visual document retrieval aims to retrieve the most relevant documents that satisfy the information need in a given textual query $q$ at the page level. The candidates for retrieval are a collection of visual document pages $\mathcal{D} = \{d_1, d_2, ..., d_{|\mathcal{D}|}\}$. Each page $d_i$ is a high-resolution screenshot containing heterogeneous content such as text, tables, figures, and layout structures.

\paragraph{SOTA Visual Document Retrievers.}
State-of-the-art visual document retrievers represent both queries and page screenshots into a multi-vector space using MLLM (e.g., Qwen-VL~\cite{bai2025qwen2} and PaliGemma~\cite{beyer2024paligemma}), and conduct late interaction. A query $q$ is encoded into a sequence of token embeddings $\mathbf{E}_q \in \mathbb{R}^{N_q \times D}$, while a page $d$ is split into patches and encoded into $\mathbf{E}_d \in \mathbb{R}^{N_d \times D}$, where $N_q$ and $N_d$ denote the number of tokens and patches, and $D$ is the embedding dimension. The relevance score is computed via late interaction using the MaxSim operator:
\begin{equation}
S(q, d) = \sum_{i=1}^{N_q} \max_{j \in [1,N_d]} \langle \mathbf{E}_q^{(i)}, \mathbf{E}_d^{(j)} \rangle,
\label{eq:rel_score}
\end{equation}
where $\langle \cdot, \cdot \rangle$ denotes the dot product, and $i$ and $j$ are the indices for the query token and document page patch, respectively. This operator first captures the strongest match for each query token, then aggregates partial evidence across all query tokens. Unlike dual-encoder methods~\cite{karpukhin2020dense} that compress inputs into single vectors and lose granularity, this approach preserves local structure and supports detailed alignment.

Training typically employs a contrastive objective using the hardest in-batch negative:
\begin{equation}
    \mathcal{L}_{\text{global}} = \log{\left(1 + \text{exp}\left(S(q, d^-) - S(q, d^+)\right)\right)},
\label{eq:match}
\end{equation}
where $d^+$ is the ground-truth relevant document page for $q$, and $d^-$ is the most similar non-relevant page in the batch. This softplus-based loss is standard in late interaction models, as it avoids computing interactions with all negatives, while still efficiently optimizing the relative ranking through in-batch hard negatives.

\section{Attention-Grounded REtriever Enhancement}

Building upon SOTA visual document retrievers, we propose the \textbf{A}ttention \textbf{G}rounded \textbf{RE}triever \textbf{E}nhancement (AGREE) training framework that leverages attention from a pretrained MLLM as auxiliary fine-grained supervision. As illustrated in Figure~\ref{fig:method}, AGREE first extracts patch-level query-to-page attention maps from an MLLM with a high-resolution grid for precise grounding. Then these attentions are downsampled to align with the retriever’s patch grid. Finally, the retriever is trained using a dual objective that combines standard global contrastive learning with fine-grained alignment to the MLLM’s attention patterns. The following subsections detail each component.

\subsection{MLLM Attention Annotation}
Given a query and its positive page, we derive proxy patch-level supervision from pre-trained MLLM through layer-wise attention extraction and multi-layer attention aggregation. 

\subsubsection{\textbf{Layer-wise Attention Extraction}} 
\label{sec:attention_annotation}
We explore two strategies to derive patch-level supervision from a pre-trained MLLM:

\textsc{Answer-Token Attention:}
Following prior work~\cite{zhuang2024promptreps, zhang2025mllms}, we extract attention from the final input token, i.e., the one immediately preceding answer generation, to capture the model’s focus before responding. Formally, given a page screenshot divided into $N_d$ patches and a prompt with $N_p$ tokens, the full input sequence has length $N_d+N_p$. We extract the cross-modal attention from the last input token, i.e., at position $N_d+N_p$ to all image patches in layer $l$ in the MLLM, yielding a query-specific attention map $\mathbf{A}^{(l)}_{\text{query}}\!\in\!\mathbb{R}^{N_d}$. Specifically, the prompt is ``\textit{Question: \{query\} Point out all content in the image that is helpful in answering the question, response in a single word or phrase.}''. 

However, raw cross-attention may highlight regions used for global information aggregation rather than semantic relevance~\cite{darcet2023vision}, to emphasize semantically meaningful attention, we normalize $\mathbf{A}^{(l)}_{\text{query}}$ using a general attention $\mathbf{A}^{(l)}_{\text{general}}$, obtained from a generic instruction: ``\textit{Write a general description of the image, response in a single word or phrase.}''.The normalized attention is:
\begin{equation}
\mathbf{A}^{(l)}
=\frac{\mathbf{A}^{(l)}_{\text{query}}}
       {\mathbf{A}^{(l)}_{\text{general}}}.
\label{eq:norm_gen}
\end{equation}
This produces an attention map that highlights regions most relevant to the query.

\textsc{Query-Token Attention:}
To avoid dependence on answer generation and instruction following capability, we also consider a direct approach. We feed the concatenated image patches and query tokens to the MLLM and extract the attention from all $N_q$ query tokens to the patches from the $l$-th layer in the MLLM. These layer-wise vectors from each token are then averaged into a unified vector:
\begin{equation}
    \mathbf{A}^{(l)} = \frac{1}{N_q} \sum_{i=1}^{N_q} \mathbf{A}^{(l,i)},
\end{equation}
where $\mathbf{A}^{(l,i)}$ is the attention vector from the $i$-th query token to all page patches in layer $l$. This method captures query-patches alignment without relying on generation behavior.

We compare the quality of the above two strategies in Section~\ref{sec:attention_quality}, and evaluate their effectiveness in guiding visual document retrieval in Section~\ref{sec:impact_of_attention}.

\subsubsection{\textbf{Multi-Layer Aggregation}} 
Given that the optimal layer for attention extraction varies with model architecture and task, and averaging attention of all layers has been shown to yield performance comparable to or better than carefully selecting a single layer~\cite{zhang2025mllms}, we integrate attention across all layers:
\begin{equation}
    \overline{\mathbf{A}} = \frac{1}{L} \sum_{l=1}^{L} \mathbf{A}^{(l)}.
\end{equation}
The resulting $\overline{\mathbf{A}} \in \mathbb{R}^{N_d}$ serves as the fine-grained, query-conditioned supervision signal, indicating the relative importance of each patch in supporting the match.

\subsection{Spatial-Preserving Attention Downsampling}

To preserve a nuanced perception of visual details and ensure high-quality supervision, we extract MLLM attention from high-resolution images, i.e., up to 2048 patches. However, this creates a resolution mismatch with the downstream retriever, which operates at lower input resolutions (e.g., 768 max patches for ColQwen, and 1024 for ColPali) due to computational constraints during training and inference. A direct application of the high-resolution attention would thus be incompatible with the retriever’s patch grid, preventing end-to-end supervision.

To bridge this gap, we employ a spatial-preserving down-sampling strategy based on adaptive max pooling. Unlike average pooling, which may dilute salient activation peaks, max pooling preserves peak attention values within each window, ensuring key regions highlighted by the MLLM remain prominent after down-sampling.

Formally, given a high-resolution attention vector $\mathbf{A}\!\in\!\mathbb{R}^{N_h}$, we first reshape it back to the 2-D grid $\mathbf{A}_{\text{high}}\!\in\!\mathbb{R}^{H_h\times W_h}$, where $H_h\times W_h=N_h$. Let the target low-resolution be $H_l\times W_l$ ($N_l=H_l\times W_l$). We obtain the aligned low-resolution map $\mathbf{A}_{\text{low}}\!\in\!\mathbb{R}^{H_l\times W_l}$ via adaptive max pooling:
\begin{equation}
\mathbf{A}_{\text{low}}[i,j]=
\max_{(u,v)\in\Omega(i,j)}
\mathbf{A}_{\text{high}}[u,v],
\label{eq:amp_pool}
\end{equation}
where $\Omega(i,j)$ defines the high-resolution region corresponding to low-resolution position $(i,j)$:
\begin{equation}
\Omega(i,j)=\Bigl\{(u,v)\;\Bigl|\;
\begin{aligned}
\bigl\lfloor\tfrac{iH_h}{H_l}\bigr\rfloor \le&\, u < \bigl\lfloor\tfrac{(i+1)H_h}{H_l}\bigr\rfloor,\\
\bigl\lfloor\tfrac{jW_h}{W_l}\bigr\rfloor \le&\, v < \bigl\lfloor\tfrac{(j+1)W_h}{W_l}\bigr\rfloor
\end{aligned}
\Bigr\}.
\end{equation}
% The resulting $\mathbf{A}_{\text{low}}$ is flattened into $\tilde{\mathbf{A}}\!\in\!\mathbb{R}^{N_l}$, forming the final fine-grained supervision signal.

This design decouples supervision quality from deployment constraints: the MLLM can operate at high resolution for accurate grounding, while the retriever can learn effectively at lower resolution, enabling scalable fine-grained knowledge transfer.

\subsection{Attention-Guided Retriever Training}
\label{sec:attention-training}

Given the high-quality, down-sampled attention, we now introduce how this fine-grained signal guides the training of the retriever. Specifically, we introduce a local alignment loss $\mathcal{L}_{\text{local}}$ applied \textbf{only to positive query-page pairs ($q, d^+$)}. This is because the MLLM’s attention patterns are meaningful only when the document is actually relevant. For irrelevant documents, no genuine alignment exists, making any attention pattern semantically ungrounded and cannot serve as valid supervision. By restricting $\mathcal{L}_{\text{local}}$ to positives, we ensure that fine-grained guidance reflects true matching logic.

\subsubsection{\textbf{Patch–Query Similarity Vector}}
\label{sec:patch-score}

To enable direct comparison between the retriever’s behavior and the MLLM’s attention, we first construct a patch-level similarity vector $\mathbf{s} \in \mathbb{R}^{N_d}$ to measure the alignment between each page region and the query. Unlike the MaxSim operator (Eq.~\ref{eq:rel_score}), which focuses on the strongest match for each query token, our goal here is to assess the overall alignment of each page patch with the entire query. Specifically, for each patch $j$ we average its similarity over all query tokens:
\begin{equation}
\label{eq:patch-sim}
s_j \;=\;
\frac{1}{N_q}\,
\sum_{i=1}^{N_q}
\bigl\langle
\mathbf{E}_{q}^{(i)},
\mathbf{E}_{d}^{(j)}
\bigr\rangle,
\qquad j=1,\dots,N_d.
\end{equation}
The resulting vector $\mathbf{s}$ captures the retriever’s holistic assessment of each patch’s relevance to the query and will be aligned with the flattened $\tilde{\mathbf{A}}=\text{flatten}(\mathbf{A}_{\text{low}})$ by the local alignment loss $\mathcal{L}_{\text{local}}$.

\subsubsection{\textbf{Fine-Grained Supervision via Attention Alignment}}
\label{sec:fg-loss-variants}

To align $\mathbf{s}$ with $\tilde{\mathbf{A}}$, we explore three strategies, covering the range from full patch-distribution matching to salient-region contrast.

\textsc{Kullback–Leibler Divergence:}
Kullback–Leibler (KL) divergence enforces strict, global alignment between the MLLM attention distribution and the retriever’s similarity scores. Specifically, we treat both as probability distributions via softmax normalization and minimize:
\begin{equation}
\label{eq:kl-loss}
\mathcal{L}_{\text{kl}}
=
\mathrm{KL}\!\bigl(
\mathrm{softmax}(\tilde{\mathbf{A}})
\;\|\;
\mathrm{softmax}(\mathbf{s})
\bigr).
\end{equation}
Since KL divergence constrains alignment across the entire image, it penalizes the model for missing even weakly attended patches. This forces the retriever to align both salient regions and low-attention areas, which can lead to overly strong constraints and increased sensitivity to noise in the attention signal.

\textsc{Top-K Salient Contrastive:}
To mitigate the impact of low-attention regions, we focus supervision on the most salient patches, encouraging the retriever to assign higher similarity scores to salient than to non-salient regions. Specifically, we sort patches in $\tilde{\mathbf{A}}$, select the top $K\%$ as the salient set $\mathcal{P}^{+}$, and treat the remainder as $\mathcal{P}^{-}$. We then apply a multi-positive contrastive loss~\cite{khosla2020supervised}:
\begin{equation}
\label{eq:intra-contrast}
\mathcal{L}_{\text{top-k}}
= - \log{\frac{\sum_{j\in\mathcal{P}^{+}}\text{exp}(s_j)}{\sum_{i\in\{\mathcal{P}^{+};\mathcal{P}^{-}\}}\text{exp}(s_i)}}.
\end{equation}
This loss promotes relative ranking between high-attention and low-attention regions. However, its effectiveness is sensitive to the choice of $K$:  a small $K$ may omit relevant content and reduce coverage, while a larger $K$ may introduce less reliable supervision by including regions with weak or noisy attention. Therefore, $K$ must be carefully tuned to balance coverage and precision.

\textsc{Cosine Similarity:}
To avoid reliance on fixed thresholds or manual tuning, we further adopt cosine similarity as a supervision objective that focuses on the alignment of score patterns rather than absolute magnitudes. It maximize the directional agreement between $\mathbf{s}$ and $\tilde{\mathbf{A}}$ using cosine similarity:
\begin{equation}
\label{eq:cos-loss}
\mathcal{L}_{\text{cos}}
=
1-
\frac{\langle\mathbf{s},\tilde{\mathbf{A}}\rangle}
{\|\mathbf{s}\|_2\;\|\tilde{\mathbf{A}}\|_2}.
\end{equation}
By discarding scale information and emphasizing vector direction, this objective naturally prioritizes alignment on the most salient regions without being penalized for minor deviations in background scores, and requiring no manual selection of thresholds. 
% Cosine similarity does not require manual selection of thresholds, making it adaptive and easy to apply across diverse query-document pairs.

We compare the effectiveness of the above constraints empirically in Section~\ref{sec:attention_loss}.

\subsubsection{\textbf{Overall Training Objective}}
\label{sec:overall-loss}
The final objective combines two objectives: global-level relevance modeling and fine-grained local rationale learning:
\begin{equation}
\label{eq:total-loss}
\mathcal{L}_{\text{total}} = \mathcal{L}_{\text{global}} + \lambda \cdot \mathcal{L}_{\text{local}},
\end{equation}
where $\mathcal{L}_{\text{global}}$ is the contrastive loss from Eq.~\ref{eq:match}, $\mathcal{L}_{\text{local}} \in \{\mathcal{L}_{\text{kl}}, \mathcal{L}_{\text{cos}},$ $\mathcal{L}_{\text{top-k}}\}$ is one of the local alignment losses, and $\lambda$ is a coefficient balances the two objectives. This dual supervision enables the retriever to learn not only whether a document is relevant, but also which regions support the match. Please note that the fine-grained supervision is used only during training. At inference time, the retriever operates independently without requiring any external attention signals or access to the MLLM.
\section{Experimental Setup}
\subsection{Datasets}

We train models on the colpali-train-set~\footnote{\url{https://huggingface.co/datasets/vidore/colpali_train_set}} introduced by ColPali~\cite{faysse2024colpali}, which contains 118,695 English-only query-page pairs sampled without hard negative mining. The dataset aggregates data from five diverse sources: DocVQA~\cite{mathew2021docvqa}, InfoVQA~\cite{mathew2022infographicvqa}, TAT-DQA~\cite{zhu2021tat}, arXivQA~\cite{li2024multimodal}, and a collection of web-scraped PDFs. Each query is associated with one labeled positive document.

For evaluation, we focus primarily on ViDoRe benchmark V2~\footnote{\url{https://huggingface.co/collections/vidore/vidore-benchmark-v2-67ae03e3924e85b36e7f53b0}}~\cite{macé2025vidorebenchmarkv2raising}, a recently introduced fully out-of-domain visual retrieval benchmark designed to reflect real-world retrieval scenarios. It emphasizes non-extractive query and long-form, multi-page, and cross-document reasoning. As shown in Table~\ref{tab:v2_statistics}, it includes diverse datasets with multiple positive pages per query. Among them, \textit{ESG Reports Human} is fully human-annotated; the other datasets use synthetic queries refined via human review.
We also report results on ViDoRe benchmark V1~\cite{faysse2024colpali}, which comprises 5 in-domain tasks derived from the same training distribution and 5 out-of-domain tasks. While it provides a generalization reference, its reliance on extractive queries makes it less challenging and not the focus of this work.

\begin{table}
\caption{Summary of ViDoRe V2 benchmark statistics.}
\vskip -0.15in
\label{tab:v2_statistics}
\begin{center}
\begin{adjustbox}{max width=1.\linewidth}
\begin{tabular}{l|ccc}
    \toprule
    Dataset & \#Queries & \#Pages & \#Pos. Pages/Query \\
    \midrule
    ESG Reports Human & 52 & 1538 & 2.46 \\
    ESG Reports & 228 & 1538 & 3.89 \\
    Economics Reports & 232 & 452 & 15.64 \\
    Biomedical Reports & 640 & 1016 & 3.22 \\
    \bottomrule
\end{tabular}
\end{adjustbox}
\end{center}
\vskip -0.15in
\end{table}

\subsection{Baselines}
We compare AGREE against a diverse set of baselines, grouped by architectural and pretraining paradigm. 

\paragraph{\textbf{Dual-Encoder Vision Language Models}}  
These models, such as \textbf{CLIP}~\cite{radford2021learning} and \textbf{SigLIP}~\cite{zhai2023sigmoid}, are pretrained on large-scale image-text pairs with a global contrastive objective. While effective for natural image retrieval, they may be limited in visual document retrieval: (1) their pretraining data focuses on natural images and captions, not complex document screenshots; (2) only global embeddings are optimized, so token-level representations may be weak and unsuitable for late interaction. We fine-tune CLIP and SigLIP following the ColPali setup, resulting in \textbf{BiCLIP} and \textbf{BiSigLIP}. We also attempt to extend them to late interaction (i.e., ColCLIP, ColSigLIP), but observe severe training instability and poor convergence, consistent with findings in~\cite{faysse2024colpali, bica2024improving}, so we exclude them from the reporting.

\paragraph{\textbf{Multi-modal Large Language Models}}
MLLMs are trained on diverse tasks. This equips them with a stronger understanding of document screenshots. More importantly, MLLMs generate high-quality, semantically meaningful token-level embeddings, making them naturally compatible with late interaction. This makes recent SOTA methods all based on MLLM. We evaluate two representative systems. \textbf{DSE}~\cite{ma2024unifying} encodes documents into a single vector using Phi-3-vision~\cite{Abdin2024Phi3TR}, representing strong single-vector retrieval. \textbf{ColPali}~\cite{faysse2024colpali} and its variant \textbf{ColQwen2.5} are SOTA late-interaction retrievers that use MaxSim scoring as we introduced in Section~\ref{sec:preliminaries}. Their backbones are PaliGemma~\cite{beyer2024paligemma} and Qwen-VL~\cite{bai2025qwen2}, respectively. While other strong models exist, such as Llama-Nemoretriever-Colembed~\cite{xu2025llamanemoretrievercolembedtopperforming} and ColNomic-Embed-Multimodal~\cite{nomicembedmultimodal2025}, their improvements stem from larger training corpora, advanced hard negative mining, and higher-resolution inputs. Since AGREE is a training strategy orthogonal to these enhancements, we adopt ColQwen2.5 as our primary baseline for a clean ablation.

\subsection{Evaluation}
The evaluation consists of three aspects: retrieval performance, fine-grained label quality, and the retriever’s similarity map interpretability.

\subsubsection{\textbf{Retrieval Performance}}
Following previous studies, we adopt \textbf{nDCG@1} and \textbf{nDCG@5} as the retrieval metrics across all tasks. These metrics are widely adopted in visual document retrieval because the goal is to support RAG, where only a few pages are used in downstream reasoning. High performance at early ranks ensures that correct information is quickly retrieved, reducing computational cost and hallucination risk. All evaluations are conducted using the official evaluation scripts\footnote{\url{https://github.com/illuin-tech/vidore-benchmark}}.

\subsubsection{\textbf{Fine-grained Label Quality}}
\label{sec:human_annotation}

\begin{figure}[t]
  \centerline{\includegraphics[width=1\linewidth]{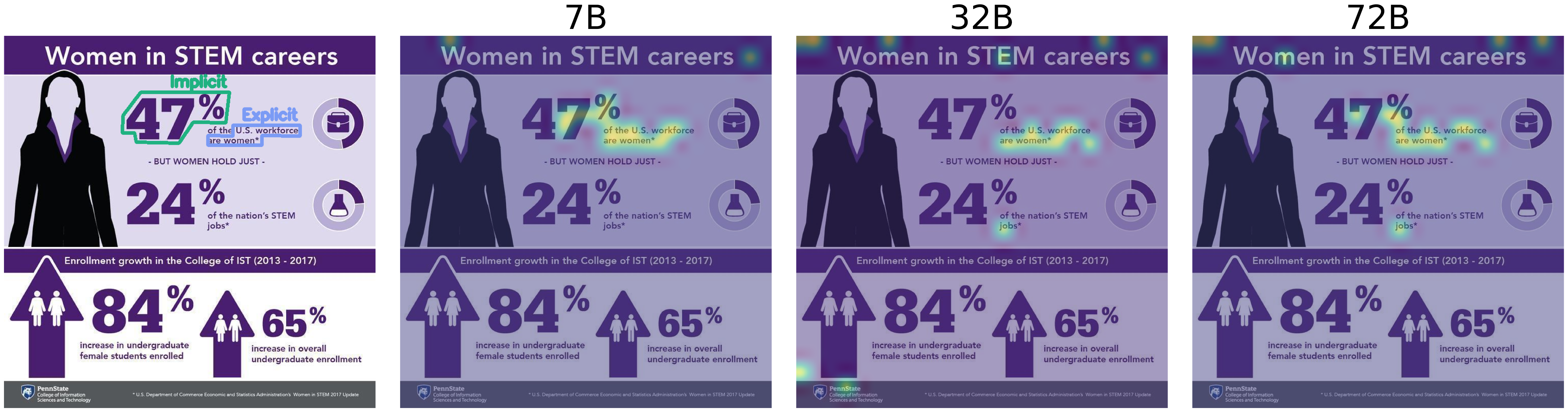}}
  \caption{Human annotation (left) and top-3\% high attention areas using ``query-token attention'' (others) given the query ``What \% of the US workforce are not women?''.}
  \label{fig:attention_of_different_model}
  \vskip -0.2in
\end{figure}

To assess the quality of generated attention, we conduct a human annotation study. We sample 250 query-document pairs, with 50 selected randomly from each of the five training data sources to ensure domain coverage. For each pair, annotators are shown the query, the positive document page, and the ground-truth answer. They are asked to draw bounding boxes around all relevant regions and classify each as either an explicit match (direct text or visual correspondence) or an implicit match (requiring inference). Two trained annotators label each instance independently (average 1.38 minutes per annotation). Inter-annotator agreement, measured by mean IoU on bounding boxes, reaches 0.66, indicating strong consistency. Discrepancies are then resolved by a senior annotator: for high-overlap cases (IoU > 0.7) with consistent labels, the clearer version is kept; for others, a refined final annotation is created. This protocol yields high-quality ground truth for the relevant area. An example is shown in Figure~\ref{fig:attention_of_different_model}. All annotators are volunteer collaborators.

% Attention quality is measured through the \textbf{coverage} of human-annotated matching regions by the top-$K$\% highest-attention areas. Higher coverage indicates better alignment between attention and human relevance judgments.
Metric Definition. To evaluate the quality of the attention maps, we measure how well the high-attention regions cover the human-annotated ground truth: the ratio of the intersection area between the top-$K$\% attention patches and the human-annotated bounding boxes to the total area of the human annotations. A higher coverage score indicates that the model's focus correctly encompasses the regions identified by humans as relevant.

\subsubsection{\textbf{Retriever’s Similarity Map Interpretability}}
\label{sec:similarity_interpretability}
Using the above human annotations, we also evaluate whether the retriever learns to highlight meaningful regions. We compute the \textbf{coverage} of human-annotated regions by the top-$K$\% highest-scoring patches in the similarity map. This metric reflects how well the retriever's behavior aligns with human judgments, providing insight into its interpretability and fine-grained reasoning capability.
 
\subsection{Implementation Details} 
To obtain fine-grained supervision from MLLM, we use \textbf{Qwen2.5-VL-7B-Instruct}~\cite{bai2025qwen25vltechnicalreport}~\footnote{\url{https://huggingface.co/Qwen/Qwen2.5-VL-7B-Instruct}} model with ``query-token attention'' strategy. This choice is motivated by its strong correlation with human relevance judgments (see Section~\ref{sec:attention_quality}). The maximum number of image patches during MLLM attention annotation is 2048. 
Our framework is implemented on two MLLM backbones, i.e., \textbf{PaliGemma-3B-pt-448}~\cite{beyer2024paligemma}~\footnote{\url{https://huggingface.co/google/paligemma-3b-pt-448}} and \textbf{Qwen2.5-VL-3B-Instruct}~\footnote{\url{https://huggingface.co/Qwen/Qwen2.5-VL-3B-Instruct}}. The number of image patches during retrieval for the two models is fixed at 1024 and a maximum of 768, respectively. To ensure a fair and comparable evaluation, we train both our method and all baselines on colpali-train-set with in-batch negatives for 3 epochs. The batch size is 128 and the learning rate is $1e-4$. We use LoRA for parameter-efficient tuning. All training can be conducted on a single H20 GPU. When multiple GPUs are available, data parallelism can be applied to further accelerate training. Importantly, the addition of the local alignment loss introduces negligible computational overhead during training. Following official recommendations, we use a batch size of 1 during inference to ensure consistency and reproducibility.

\section{Overall Performance}

\begin{table*}[!ht]
\caption{Test results on ViDoRe benchmark V2. The best results are bolded, and the suboptimal result is underlined. ${\dagger}$ and $*$ denote significant improvement over the baseline using the same backbone.}
\vskip -0.15in
\label{tab:overall_result_v2}
\begin{center}
\setlength{\tabcolsep}{3pt}
% \begin{adjustbox}{max width=1.\linewidth}
\begin{tabular}{l|cccccccc|cc}
    \toprule
     & \multicolumn{2}{c}{Esg} & \multicolumn{2}{c}{Biomedical} & \multicolumn{2}{c}{Economics} & \multicolumn{2}{c|}{EsgHuman} & \multicolumn{2}{c}{Avg} \\
    Model & nDCG@1 & nDCG@5 & nDCG@1 & nDCG@5 & nDCG@1 & nDCG@5 & nDCG@1 & nDCG@5 & nDCG@1 & nDCG@5 \\
    \midrule
    \multicolumn{11}{l}{\emph{Vision-Language Models}} \\
    % CLIP (0-shot) & 6.58 & 10.45 & 15.94 & 18.40 & 17.24 & 15.37 & 13.46 & 18.86 & 13.30 &	15.77 \\
    BiCLIP & 10.53 & 11.38 & 20.16 & 22.39 & 26.29 & 22.83 & 	17.31 &	22.74 &	18.57 &	19.84 \\
    % ColCLIP & 0.0 & 0.0	& 0.0 & 0.0 & 6.9 & 5.0 & 0.0 & 0.5 &	1.7 & 1.4\\
    % ColFineCLIP & 0.0 & 0.0	& 0.0 & 0.0 & 6.9 & 5.0 & 0.0 & 0.5 &	1.7 & 1.4\\
    % ColFGCLIP & 0.0 & 0.0	& 0.0 & 0.0 & 6.9 & 5.0 & 0.0 & 0.5 &	1.7 & 1.4\\
    SigLIP (0-shot) & 26.75 & 28.19 & 20.16 & 24.30 & 18.97 & 16.34 & 19.23	& 23.07 &	21.28 &	22.98 \\
    BiSigLIP & 29.83 & 34.92 & 25.78 & 29.05 & 30.60 & 28.23 & 34.62 & 45.83 & 30.21 & 34.51 \\
    % ColSigLIP & 0.0	& 0.3 &	0.0	& 0.20	& 3.0 & 3.1 & 0.0 & 0.0	& 0.8 & 0.9\\
    \midrule
    \multicolumn{11}{l}{\emph{Multi-modal Large Language Models}} \\
    DSE & 47.37 & 55.15 & 52.66 & 55.29 & 62.50 & 52.80 & 57.69 & 61.74 & 55.06 & 56.25 \\
    ColPali & 51.75 & 54.05 & 52.03 & 57.56 & 55.17 & \underline{53.30} & 54.49 & 63.67 & 53.36 & 57.14 \\
    % ColQwen2.5 & 52.2 & 56.2 & \textbf{56.4} & \textbf{60.6} & 53.4 & 52.4 & 60.3 & \underline{65.7} & 55.6 & \underline{58.7} \\
    ColQwen2.5 & 51.75 & 57.12 & \textbf{56.56} & \textbf{60.64} & 52.59 & 52.90 & 58.33& 63.70 & 54.81 & \underline{58.59} \\
    \midrule
    \multicolumn{11}{l}{\emph{Ours}} \\
    AGREEPali & \underline{57.46$^{\dagger}$} & \textbf{58.87$^{\dagger}$} & 52.03 & 55.51 & \textbf{59.05$^{\dagger}$} & 51.33 & \underline{67.95$^{\dagger}$} & 65.28$^{\dagger}$ & \underline{59.12} & 57.75\\
    % AGREEQwen2.5 & \textbf{54.4} & \textbf{58.9} & 55 & 60.2 & \textbf{62.9} & \textbf{59.3} & \textbf{60.3} & \textbf{66.5} & \textbf{58.1} & \textbf{61.2} \\
    % AGREEQwen2.5 (72b+at) & 62.28 & 60.91 & 57.66 & 60.29 & 58.62 & 55.58 & 63.14 & 66.25 & 60.43 & 60.76 \\
    AGREEQwen2.5 & \textbf{60.09$^{*}$} & \underline{58.43} & \underline{56.41} & \underline{60.49} & \textbf{59.05$^{*}$} & \textbf{56.50$^{*}$} & \textbf{71.80$^{*}$} & \textbf{70.73$^{*}$} & \textbf{61.84} & \textbf{61.54} \\
    \bottomrule
\end{tabular}
% \end{adjustbox}
\end{center}
\vskip -0.1in
\end{table*}

\begin{figure}[t]
  \centering
  \begin{subfigure}[t]{0.43\linewidth}
    \centering
    \includegraphics[width=\linewidth]{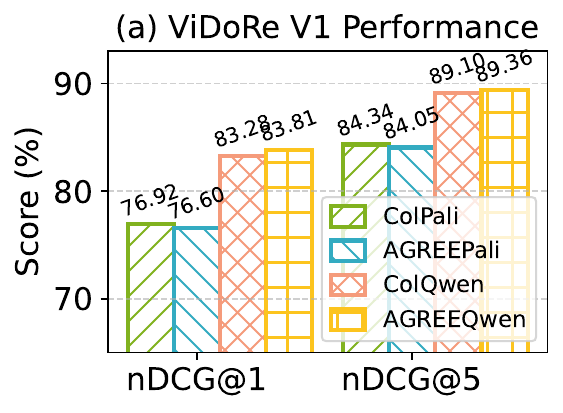}
    \label{fig:v1_results}
  \end{subfigure}
  \hfill
  \begin{subfigure}[t]{0.55\linewidth}
    \centering
    \includegraphics[width=\linewidth]{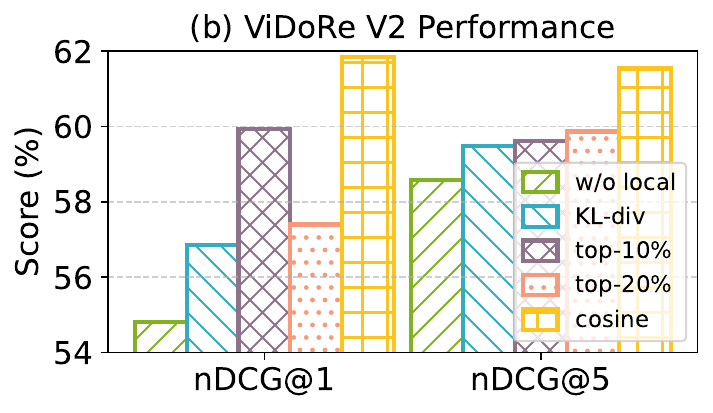}
    \label{fig:global_loss}
  \end{subfigure}
  \vskip -0.15in
  \caption{(a) Results on ViDoRe V1 using PaliGemma and Qwen2.5-VL as backbones; (b) Results on ViDoRe V2 with different local alignment losses $\mathcal{L}_{\text{local}}$.}
  \label{fig:results}
  \vskip -0.1in
\end{figure}

% The overall results on the ViDoRe V2 benchmark are summarized in Table~\ref{tab:overall_result_v2}. Specifically, AGREE achieves substantial improvements: AGREEQwen2.5 achieves 61.84\% on average nDCG@1 and 61.54\% on average nDCG@5, obtains +7.03\% and +2.95\% absolute gains compared with the strongest baseline ColQwen2.5, highlighting that AGREE is effective in challenging, realistic retrieval scenarios. By providing signals for identifying matching regions, AGREE effectively guides the model to discover non-trivial, fine-grained alignments that are poorly captured by binary relevance signals alone. Further comparisons between ColPali vs. AGREEPali show that AGREE exhibits consistent performance improvements, verifying its strong generalization and plug-and-play applicability across diverse backbones. 

The overall results on the ViDoRe V2 benchmark are summarized in Table~\ref{tab:overall_result_v2}. Specifically, AGREE achieves substantial improvements: AGREEQwen2.5 achieves 61.84\% on average nDCG@1 and 61.54\% on average nDCG@5, obtains +7.03\% and +2.95\% absolute gains compared with the strongest baseline ColQwen2.5. Remarkably, despite being trained on the extractive Copali dataset, AGREE excels on this reasoning-heavy benchmark. This indicates that the framework successfully distills the intrinsic reasoning capabilities of the teacher MLLM. By mimicking the teacher's semantic attention rather than relying on surface-level training signals, the retriever learns a robust alignment mechanism that generalizes effectively to implicit, non-extractive queries.

Figure~\ref{fig:results} (a) presents results on ViDoRe V1, showing that AGREE maintains competitive performance on simpler, explicit tasks. This indicates that for explicit matches, global-level relevance labels are often sufficient to guide effective retrieval. However, crucially, the incorporation of fine-grained supervision does not negatively impact performance on these explicit tasks, confirming that AGREE enhances semantic reasoning without compromising the model's ability to handle basic surface-level matching.

\section{Further Analysis}
\subsection{MLLM Attention Annotation Evaluation}
\label{sec:attention_quality}

\begin{figure}[t]
  \centerline{\includegraphics[width=1\linewidth]{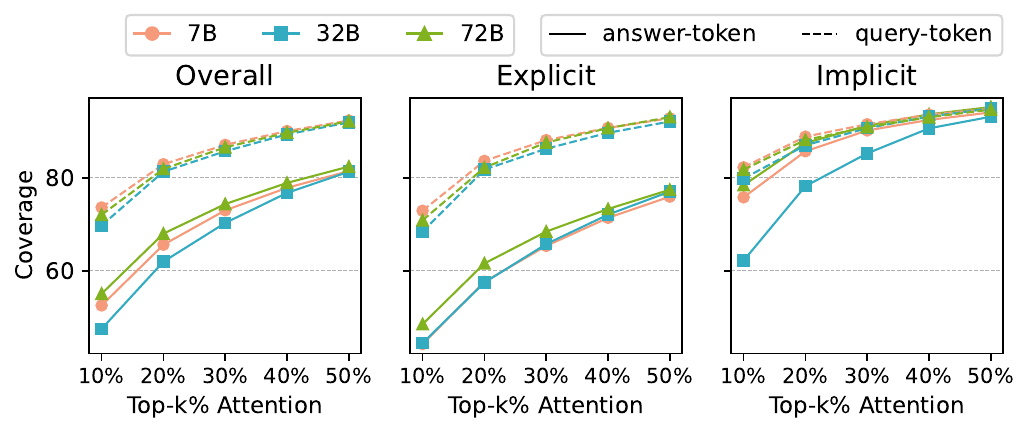}}
  % \vskip -0.15in
  \caption{Coverage of human-annotated matching areas by top-k\% attention regions with models of different sizes and different attention extraction strategies.}
  \label{fig:attention_coverage_iou}
  % \vskip -0.1in
\end{figure}

To measure supervision quality, we evaluate different MLLM-derived attention maps using the human annotations introduced in Section~\ref{sec:human_annotation}. We compare two strategies introduced in Section~\ref{sec:attention_annotation}, i.e., \emph{answer-token attention} and \emph{query-token attention}, across Qwen2.5-VL models of varying sizes.

Results in Figure~\ref{fig:attention_coverage_iou} indicate that \textbf{the direct ``query-token attention'' achieves the highest alignment with human judgments, consistently covering both explicit and implicit matches.} When comparing model sizes, we find that even a 7B model achieves strong performance under such a paradigm. As visualized in Figure~\ref{fig:attention_of_different_model}, the top-3\% attention regions across model sizes exhibit strong alignment and generally cover both explicitly mentioned and contextually implied content, suggesting robust and size-invariant localization ability. In contrast, the ``answer-token strategy'' relies heavily on instruction-following and reasoning capabilities, which are more developed in larger models, leading to a significant performance gap. We also observe that the 32B model performs poorly, likely due to excessive chain-of-thought reasoning, even when such reasoning is disabled. In addition, attentions based on the ``answer-token strategy'' mainly pay more on the implicit areas, leading to more serious omissions of the explicit matching. These findings motivate our default choice to use the \textbf{``query-token attention'' from Qwen2.5-VL-7B-Instruct} as fine-grained supervision labels.

\subsection{Impact of Attention Quality on Performance}

\label{sec:impact_of_attention}
\begin{table}
\caption{Average test results based on attentions with different models and extraction strategy.}
% \vskip -0.15in
\label{tab:attention_results}
\setlength{\tabcolsep}{4pt}
\begin{center}
\begin{tabular}{ll|cc|cc}
    \toprule
    & & \multicolumn{2}{c|}{ViDoRe-V1} & \multicolumn{2}{c}{ViDoRe-V2} \\
    Size & Strategy & nDCG@1 & nDCG@5 & nDCG@1 & nDCG@5 \\
    \midrule
    - & - & 83.28 & 89.10 & 54.81 & 58.59 \\
    7B & answer & 83.15 & 88.99 & 57.10 & 59.14\\
    72B & answer & 83.23 & 88.89 & 60.43 & 60.76 \\
    7B & query & 83.81 & 89.36 & 61.84 & 61.54 \\
    \bottomrule
\end{tabular}
\end{center}
% \vskip -0.1in
\end{table}

% A critical question is how the choice of teacher model and attention extraction strategy affects retrieval performance. Different MLLMs or extraction methods may yield qualitatively different attention patterns. We analyze this by comparing the performance of retrievers trained with different attention variants. As shown in Table~\ref{tab:attention_results}, \textbf{the effectiveness of attention-guided training is strongly correlated with its alignment to human judgments.} The 7B model with query-token attention, which best aligns with human annotations, achieves the largest gains on both V1 and V2. In contrast, answer-token variants yield smaller gains on V2, and even negative gains on V1. This suggests that while it can capture implicit matches, its weak alignment with explicit content harms performance in extractive settings. This highlights that \textbf{effective fine-grained supervision must cover both surface-level and semantic matches to generalize across tasks.} This analysis also reveals that AGREE’s success hinges on the quality of the proxy labels. As future methods produce more accurate labels, AGREE can directly leverage them for further gains.

A critical question is how the choice of MLLMs and attention extraction strategy affects retrieval performance, as different configurations yield qualitatively different attention patterns. We analyze this by comparing retrievers trained with various attention variants in Table~\ref{tab:attention_results}. The results demonstrate that \textbf{the effectiveness of AGREE is strongly correlated with the attention's alignment to human judgments.} The ``7B query-token" strategy, which aligns best with human annotations (Section~\ref{sec:attention_quality}), achieves the largest gains on both benchmarks. Conversely, ``answer-token" variants show weaker alignment with explicit content, which limits their effectiveness. This results in suboptimal gains on V2 and even negative impacts on the extractive V1 benchmark. Notably, \textbf{superior extraction strategies outweigh model scaling.} The 7B model using the "query-token" strategy significantly outperforms the much larger 72B model using the "answer-token" strategy. This implies that simply scaling up the teacher model is less effective than adopting a precise attention extraction mechanism, and effective fine-grained supervision must cover both surface-level and semantic matches to generalize across tasks. This analysis also reveals that AGREE’s success hinges on the quality of the proxy labels. As future methods produce more accurate labels, AGREE can directly leverage them for further gains.

\subsection{Effect of Attention Loss Weighting}
\begin{table}
\caption{Average test results with different coefficient $\lambda$.}
% \vskip -0.15in
\label{tab:lambda_results}
\begin{center}
% \begin{adjustbox}{max width=1.\linewidth}
\begin{tabular}{l|cc|cc}
    \toprule
    & \multicolumn{2}{c|}{ViDoRe-V1} & \multicolumn{2}{c}{ViDoRe-V2} \\
    $\lambda$ & nDCG@1 & nDCG@5 & nDCG@1 & nDCG@5 \\
    \midrule
    0 & 83.28 & 89.10 & 54.81 & 58.59 \\
    5e-1 & 82.82 & 89.20 & 56.71 & 58.28\\
    1e-1 & 83.81 & 89.36 & 61.84 & 61.54 \\
    5e-2 & 83.29 & 89.21 & 58.71 & 59.94 \\
    \bottomrule
\end{tabular}
% \end{adjustbox}
\end{center}
% \vskip -0.1in
\end{table}

We evaluate the impact of the attention loss coefficient $\lambda$. As shown in Table~\ref{tab:lambda_results}, retrieval performance achieves its best at $\lambda=1e-1$. This indicates that fine-grained supervision must be balanced. \textbf{Considering that MLLM's attention inevitably contains noise, a strong attention objective risks overfitting to noisy attention from the MLLM. While insufficient guidance leaves the fine-grained cues underutilized.} On ViDoRe V1, performance remains stable across $\lambda$ values, consistent with its reliance on strong, explicit matching signals. This supports our earlier hypothesis: auxiliary supervision has diminishing returns when the primary task signal is already sufficient

\subsection{Comparison of Local Alignment Loss}
\label{sec:attention_loss}

We compare the local alignment losses described in Section~\ref{sec:fg-loss-variants}, including full-distribution matching (KL-divergence), salient-region contrast (top-K), and directional alignment (cosine). As shown in Figure~\ref{fig:results} (b), the cosine loss achieves the best performance. The superiority of cosine loss indicates that \textbf{effective guidance should emphasize directional agreement on salient regions, rather than enforce full distribution matching.} Cosine loss naturally focuses on high-attention patches without penalizing minor deviations in low-salience areas, making it robust and effective. Top-K performance is sensitive to $K$, trading off coverage and noise. KL-divergence performs worst because it enforces full distribution matching, forcing the model to fit even irrelevant patches.

\subsection{Selective Attention Supervision}

\begin{figure}[t]
  \centerline{\includegraphics[width=1.\linewidth]{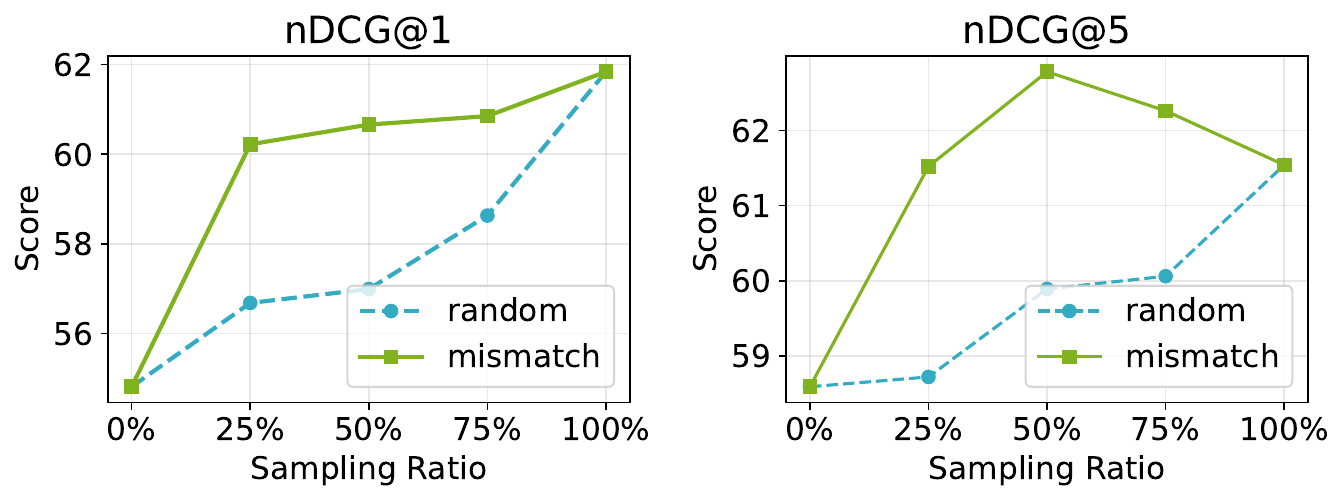}}
  % \vskip -0.15in
  \caption{Retrieval performance on ViDoRe V2 related to the proportion of samples receiving attention guidance, under random and mismatch-first sampling strategies.}
  \label{fig:different_sample_score}
  % \vskip -0.2in
\end{figure}

We evaluate AGREE’s data efficiency by applying attention supervision to subsets of the training data using two strategies: (1) \emph{random sampling}, and (2) \emph{mismatch-first sampling}. The latter acts as a hard sample mining strategy, prioritizing instances where the retriever’s similarity map diverges most from the MLLM’s attention. As shown in Figure~\ref{fig:different_sample_score}, while random sampling yields steady gains as more data is labeled, mismatch-based sampling is significantly more efficient. Applying supervision to just the top-25\% most misaligned samples yields a +5.41\% absolute gain in average nDCG@1, and further supervision brings steady but diminishing returns. This shows that \textbf{attention guidance is most critical for hard instances where the retriever's reasoning is flawed. Focusing on these samples can significantly reduce annotation costs while preserving performance.} 
% It is important to note that our mismatch-first sampling is a practical compromise, as it requires computing attention for all samples to identify hard instances. While mining hard instances via retrieval performance is a common alternative, it is unreliable here due to the dataset's incomplete labels: low-performing instances contain a large number of unlabeled positives, i.e., false negatives. Focusing on them introduces noise rather than useful supervision. In contrast, the mismatch between MLLM attention and retriever similarity better reflects model uncertainty. Importantly, this limitation is not inherent to AGREE. With more accurate annotated data, retrieval-based hard instance mining could enable true attention labeling cost reduction.
However, we acknowledge that this label efficiency is currently theoretical. Identifying these ``mismatch-based" hard samples requires computing attention for the entire dataset first. Ideally, one would mine hard samples using retrieval metrics without needing attention labels. Unfortunately, this is infeasible with the current training set due to incomplete labeling: low-performing instances contain a large number of unlabeled positives, i.e., false negatives. Focusing on them introduces noise rather than useful supervision. Therefore, we use the attention-similarity mismatch as the reliable proxy for difficulty. This analysis demonstrates the theoretical potential for data efficiency: with more accurate annotated data, retrieval-based hard instance mining could enable true attention labeling cost reduction.

% We evaluate AGREE’s performance when applying attention supervision with different proportions of training data, under two sampling strategies: (1) \emph{random sampling}, where samples are selected uniformly, and (2) \emph{mismatch-first sampling}, where we prioritize samples with the largest mismatch between similarity scores and attention pattern. As shown in Figure~\ref{fig:different_sample_score_fix_seed}, when using the random strategy, performance improves gradually as more samples receive guidance, reaching the best performance when applied to all data. In contrast, the mismatch-first strategy achieves nearly the same gain with far fewer samples: adding guidance to just the top-25\% most misaligned instances yields a +5.1 point jump in nDCG@1, and further prioritization brings steady improvements with diminishing returns. This fine-grained attention guidance is not equally beneficial across all samples; the impact is concentrated on challenging instances where alignment is non-trivial. Therefore, when training resources are limited, rather than uniformly applying costly distillation, one can prioritize samples where the model’s current understanding diverges from the teacher’s grounding, achieving comparable gains with significantly less computational overhead.

\begin{figure}[t]
  \centerline{\includegraphics[width=1\linewidth]{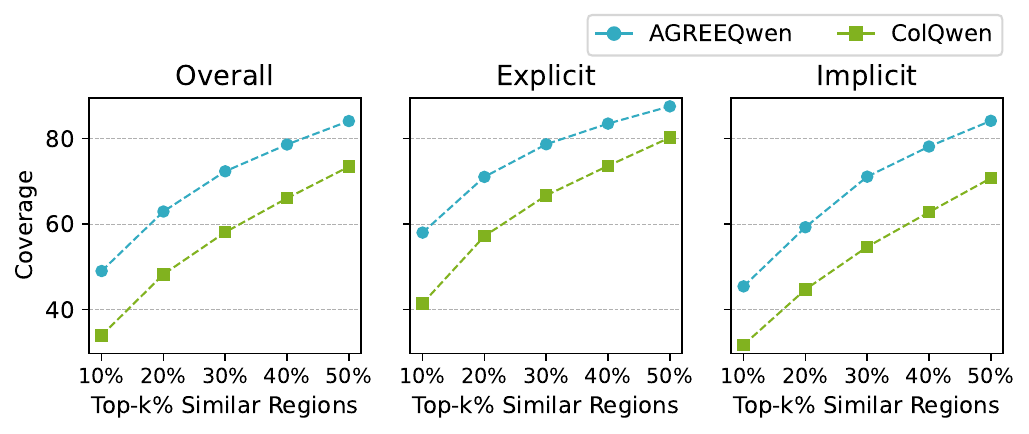}}
  \caption{Coverage of human-annotated matching areas by ColQwen2.5 and ours AGREEQwen2.5.}
  \label{fig:attention_coverage_change}
  % \vskip -0.15in
\end{figure}

\begin{figure*}[t]
  \centerline{\includegraphics[width=1\linewidth]{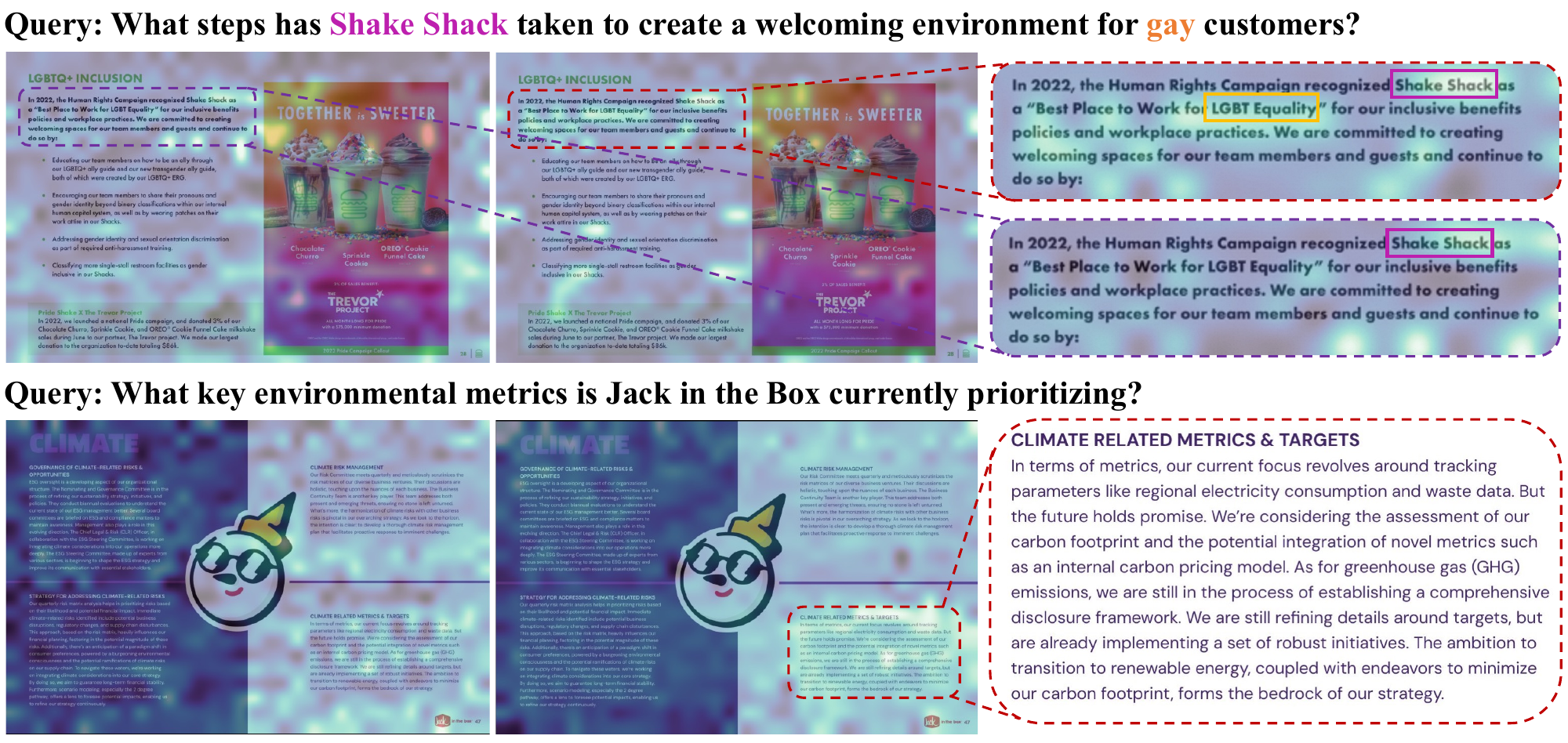}}
  \caption{Late interaction heatmaps of ColQwen2.5 (left) and AGREEQwen2.5 (middle). The right panels zoom in on key regions.}
  \label{fig:case}
  % \vskip -0.15in
\end{figure*}

\subsection{Fine-Grained Alignment Behavior Analysis}

Beyond retrieval metrics, we evaluate the plausibility of the retrievers' similarity maps using the human annotations as described in Section~\ref{sec:similarity_interpretability}. As shown in Figure~\ref{fig:attention_coverage_change}, AGREEQwen2.5 achieves significantly higher coverage of human-annotated regions than ColQwen2.5, with consistent improvement on both explicit and implicit matches. This suggests that the retrieval performance improvement introduced by AGREE is rooted in its ability to reshape how a retriever computes relevance. \textbf{By combining local fine-grained grounding signals during training, the retriever learns not only whether a document matches, but also which regions contribute to the match, moving toward more interpretable, rationale-aware matching.}

\subsection{Case Study}

% The case study further reveals a key insight: document-level 0/1 relevance labels are insufficient for learning complex matching patterns, especially those involving lexical variation or implicit inference, and fine-grained attention guidance is critical to bridge this gap. 
The case study intuitively demonstrates the limitations of global-level relevance labels and the value of fine-grained guidance. Figure~\ref{fig:case} visualizes the patch-level similarity maps~\footnote{\url{https://github.com/illuin-tech/colpali/tree/main}} of two examples from the ESG Restaurant Humans task in ViDoRe V2, where queries require synonym understanding and implicit inference.

In the first example, the query uses ``gay'' while the document uses ``LGBT''. The baseline highlights only the exact match on ``Shake Shack''. AGREEQwen2.5, however, highlights both the ``Shake Shack'' and the ``LGBT''. This indicates that \textbf{AGREE successfully helps capture the semantic equivalence between lexical variation.} 
In the second example, besides the keyword ``metrics'', more matches exist in the answer area, i.e., the bottom-right corner. This matching pattern is non-trivial, and the baseline model assigns scattered scores across the page, failing to identify the relevant region. In contrast, AGREEQwen2.5 highlights the answer region successfully, demonstrating that \textbf{AGREE helps the retriever learn to localize complex, inferential matches.} 

These behaviors highlight that global-level relevance signals are insufficient for learning nuanced matching patterns. Retrievers trained only with coarse supervision tend to rely on surface-level matching and fail to capture semantically meaningful alignments. In contrast, AGREE can effectively help the model capture both explicit and implicit matches by providing more detailed guidance.

\section{Conclusion and Future Work}

In this work, we introduced AGREE, a novel training framework that enhances visual document retrieval through fine-grained supervision. We demonstrated that retrievers trained with only global relevance labels often resorted to surface-level cues and failed to capture implicit semantic connections, resulting in suboptimal performance on queries requiring implicit reasoning or non-extractive alignment. To alleviate this limitation, we presented AGREE, which goes beyond global relevance signals by incorporating local-level supervision derived from the attention maps of pre-treined MLLMs. This supervision required no manual annotation, making the framework scalable. We demonstrated the effectiveness of AGREE through significant performance improvements on the challenging ViDoRe V2 benchmark. Ablation studies confirmed the importance of high-quality attention signals, careful local-loss design, and balanced global-local supervision. Further analysis revealed that AGREE shifts retrieval behavior from surface-level keyword matching toward semantically grounded, rationale-aware alignment. In the future, more precise forms of grounding, such as object-detection style annotations, could further improve fine-grained alignment, offering a promising direction for building even more interpretable visual document retrieval systems.

\section*{Acknowledgement}
% This work was funded by the National Natural Science Foundation of China (NSFC) under Grants No. 62302486, the Innovation Project of ICT CAS under Grants No. E361140, the CAS Special Research Assistant Funding Project, the Lenovo-CAS Joint Lab Youth Scientist Project, the project under Grants No. JCKY2022130C039, and the Strategic Priority Research Program of the CAS under Grants No. XDB0680102.
This work was supported by Alibaba Group through Alibaba Research Intern Program.
This work was funded by the National Natural Science Foundation of China (NSFC) under Grant No. 62302486 and the Innovation Project of ICT CAS under Grant No. E361140.

% \appendix
% \input{appendix/appendix}

% \newpage
% \section{Ethics Statement}

% This work is based exclusively on publicly available datasets and open-source models. No personal or sensitive information was involved. As such, the study does not raise any notable ethical concerns. In commitment to open science and reproducibility, we will release all code used in this research. This transparency enables independent verification of our results and promotes the broader adoption and extension of our methods.
\bibliographystyle{ACM-Reference-Format}
\bibliography{sigir}

@String{Computer = "{IEEE} Computer" }

@String{Springer = "Springer-Verlag" }

@inproceedings{faysse2024colpali,
  title={Colpali: Efficient document retrieval with vision language models},
  author={Faysse, Manuel and Sibille, Hugues and Wu, Tony and Omrani, Bilel and Viaud, Gautier and Hudelot, C{\'e}line and Colombo, Pierre},
  booktitle={The Thirteenth International Conference on Learning Representations},
  year={2024}
}

@misc{macé2025vidorebenchmarkv2raising,
      title={ViDoRe Benchmark V2: Raising the Bar for Visual Retrieval}, 
      author={Quentin Macé and António Loison and Manuel Faysse},
      year={2025},
      eprint={2505.17166},
      archivePrefix={arXiv},
      primaryClass={cs.IR},
      url={https://arxiv.org/abs/2505.17166}, 
}

@misc{nomicembedmultimodal2025,
  title={Nomic Embed Multimodal: Interleaved Text, Image, and Screenshots for Visual Document Retrieval},
  author={Nomic Team},
  year={2025},
  publisher={Nomic AI},
  url={https://nomic.ai/blog/posts/nomic-embed-multimodal},
}

@inproceedings{khattab2020colbert,
  title={Colbert: Efficient and effective passage search via contextualized late interaction over bert},
  author={Khattab, Omar and Zaharia, Matei},
  booktitle={Proceedings of the 43rd International ACM SIGIR conference on research and development in Information Retrieval},
  pages={39--48},
  year={2020}
}

@inproceedings{ma2024unifying,
  title={Unifying Multimodal Retrieval via Document Screenshot Embedding},
  author={Ma, Xueguang and Lin, Sheng-Chieh and Li, Minghan and Chen, Wenhu and Lin, Jimmy},
  booktitle={Proceedings of the 2024 Conference on Empirical Methods in Natural Language Processing},
  pages={6492--6505},
  year={2024}
}

@article{zhang2025mllms,
  title={Mllms know where to look: Training-free perception of small visual details with multimodal llms},
  author={Zhang, Jiarui and Khayatkhoei, Mahyar and Chhikara, Prateek and Ilievski, Filip},
  journal={arXiv preprint arXiv:2502.17422},
  year={2025}
}

@article{wang2025exploring,
  title={Exploring Implicit Visual Misunderstandings in Multimodal Large Language Models through Attention Analysis},
  author={Wang, Pengfei and Xu, Guohai and Wang, Weinong and Yang, Junjie and Lou, Jie and Xue, Yunhua},
  journal={arXiv preprint arXiv:2505.10541},
  year={2025}
}

@misc{bai2025qwen25vltechnicalreport,
      title={Qwen2.5-VL Technical Report}, 
      author={Shuai Bai and Keqin Chen and Xuejing Liu and Jialin Wang and Wenbin Ge and Sibo Song and Kai Dang and Peng Wang and Shijie Wang and Jun Tang and Humen Zhong and Yuanzhi Zhu and Mingkun Yang and Zhaohai Li and Jianqiang Wan and Pengfei Wang and Wei Ding and Zheren Fu and Yiheng Xu and Jiabo Ye and Xi Zhang and Tianbao Xie and Zesen Cheng and Hang Zhang and Zhibo Yang and Haiyang Xu and Junyang Lin},
      year={2025},
      eprint={2502.13923},
      archivePrefix={arXiv},
      primaryClass={cs.CV},
      url={https://arxiv.org/abs/2502.13923}, 
}

@misc{liu2024llavanext,
    title={LLaVA-NeXT: Improved reasoning, OCR, and world knowledge},
    url={https://llava-vl.github.io/blog/2024-01-30-llava-next/},
    author={Liu, Haotian and Li, Chunyuan and Li, Yuheng and Li, Bo and Zhang, Yuanhan and Shen, Sheng and Lee, Yong Jae},
    month={January},
    year={2024}
}

@misc{liu2023llava,
      title={Visual Instruction Tuning}, 
      author={Liu, Haotian and Li, Chunyuan and Wu, Qingyang and Lee, Yong Jae},
      publisher={NeurIPS},
      year={2023},
}

@article{lewis2020retrieval,
  title={Retrieval-augmented generation for knowledge-intensive nlp tasks},
  author={Lewis, Patrick and Perez, Ethan and Piktus, Aleksandra and Petroni, Fabio and Karpukhin, Vladimir and Goyal, Naman and K{\"u}ttler, Heinrich and Lewis, Mike and Yih, Wen-tau and Rockt{\"a}schel, Tim and others},
  journal={Advances in neural information processing systems},
  volume={33},
  pages={9459--9474},
  year={2020}
}

@article{sparck1972statistical,
  title={A statistical interpretation of term specificity and its application in retrieval},
  author={Sparck Jones, Karen},
  journal={Journal of documentation},
  volume={28},
  number={1},
  pages={11--21},
  year={1972},
  publisher={MCB UP Ltd}
}

@article{robertson2009probabilistic,
  title={The probabilistic relevance framework: BM25 and beyond},
  author={Robertson, Stephen and Zaragoza, Hugo and others},
  journal={Foundations and Trends{\textregistered} in Information Retrieval},
  volume={3},
  number={4},
  pages={333--389},
  year={2009},
  publisher={Now Publishers, Inc.}
}

@article{reimers2019sentence,
  title={Sentence-bert: Sentence embeddings using siamese bert-networks},
  author={Reimers, Nils and Gurevych, Iryna},
  journal={arXiv preprint arXiv:1908.10084},
  year={2019}
}

@inproceedings{karpukhin2020dense,
  title={Dense Passage Retrieval for Open-Domain Question Answering.},
  author={Karpukhin, Vladimir and Oguz, Barlas and Min, Sewon and Lewis, Patrick SH and Wu, Ledell and Edunov, Sergey and Chen, Danqi and Yih, Wen-tau},
  booktitle={EMNLP (1)},
  pages={6769--6781},
  year={2020}
}

@article{wang2022text,
  title={Text embeddings by weakly-supervised contrastive pre-training},
  author={Wang, Liang and Yang, Nan and Huang, Xiaolong and Jiao, Binxing and Yang, Linjun and Jiang, Daxin and Majumder, Rangan and Wei, Furu},
  journal={arXiv preprint arXiv:2212.03533},
  year={2022}
}

@article{memon2020handwritten,
  title={Handwritten optical character recognition (OCR): A comprehensive systematic literature review (SLR)},
  author={Memon, Jamshed and Sami, Maira and Khan, Rizwan Ahmed and Uddin, Mueen},
  journal={IEEE access},
  volume={8},
  pages={142642--142668},
  year={2020},
  publisher={IEEE}
}

@inproceedings{smith2007overview,
  title={An overview of the Tesseract OCR engine},
  author={Smith, Ray},
  booktitle={Ninth international conference on document analysis and recognition (ICDAR 2007)},
  volume={2},
  pages={629--633},
  year={2007},
  organization={IEEE}
}

@article{yasunaga2022retrieval,
  title={Retrieval-augmented multimodal language modeling},
  author={Yasunaga, Michihiro and Aghajanyan, Armen and Shi, Weijia and James, Rich and Leskovec, Jure and Liang, Percy and Lewis, Mike and Zettlemoyer, Luke and Yih, Wen-tau},
  journal={arXiv preprint arXiv:2211.12561},
  year={2022}
}

@inproceedings{chen2024vtqa,
  title={VTQA: Visual Text Question Answering via Entity Alignment and Cross-Media Reasoning},
  author={Chen, Kang and Wu, Xiangqian},
  booktitle={Proceedings of the IEEE/CVF Conference on Computer Vision and Pattern Recognition},
  pages={27218--27227},
  year={2024}
}

@inproceedings{wei2024uniir,
  title={Uniir: Training and benchmarking universal multimodal information retrievers},
  author={Wei, Cong and Chen, Yang and Chen, Haonan and Hu, Hexiang and Zhang, Ge and Fu, Jie and Ritter, Alan and Chen, Wenhu},
  booktitle={European Conference on Computer Vision},
  pages={387--404},
  year={2024},
  organization={Springer}
}

@article{jiang2024vlm2vec,
  title={Vlm2vec: Training vision-language models for massive multimodal embedding tasks},
  author={Jiang, Ziyan and Meng, Rui and Yang, Xinyi and Yavuz, Semih and Zhou, Yingbo and Chen, Wenhu},
  journal={arXiv preprint arXiv:2410.05160},
  year={2024}
}

@inproceedings{tanaka2023slidevqa,
  title={Slidevqa: A dataset for document visual question answering on multiple images},
  author={Tanaka, Ryota and Nishida, Kyosuke and Nishida, Kosuke and Hasegawa, Taku and Saito, Itsumi and Saito, Kuniko},
  booktitle={Proceedings of the AAAI Conference on Artificial Intelligence},
  volume={37},
  number={11},
  pages={13636--13645},
  year={2023}
}

@misc{pymupdf,
    title = {PyMuPDF},
    author = {pymupdf},
    year = {2012},
    howpublished = {\url{https://github.com/pymupdf/PyMuPDF}}
}

@misc{pdfminer,
    title = {pdfminer.six},
    author = {pdfminer},
    year = {2014},
    howpublished = {\url{https://github.com/pdfminer/pdfminer.six}}
}

@inproceedings{luo2023unifying,
  title={Unifying text, tables, and images for multimodal question answering},
  author={Luo, Haohao and Shen, Ying and Deng, Yang},
  year={2023},
  organization={Association for Computational Linguistics}
}

@article{liu2023mmhqa,
  title={Mmhqa-icl: Multimodal in-context learning for hybrid question answering over text, tables and images},
  author={Liu, Weihao and Lei, Fangyu and Luo, Tongxu and Lei, Jiahe and He, Shizhu and Zhao, Jun and Liu, Kang},
  journal={arXiv preprint arXiv:2309.04790},
  year={2023}
}

@inproceedings{yu2023unified,
  title={Unified Language Representation for Question Answering over Text, Tables, and Images},
  author={Yu, Bowen and Fu, Cheng and Yu, Haiyang and Huang, Fei and Li, Yongbin},
  booktitle={Findings of the Association for Computational Linguistics: ACL 2023},
  pages={4756--4765},
  year={2023}
}

@article{cho2024m3docrag,
  title={M3docrag: Multi-modal retrieval is what you need for multi-page multi-document understanding},
  author={Cho, Jaemin and Mahata, Debanjan and Irsoy, Ozan and He, Yujie and Bansal, Mohit},
  journal={arXiv preprint arXiv:2411.04952},
  year={2024}
}

@article{hinton2015distilling,
  title={Distilling the knowledge in a neural network},
  author={Hinton, Geoffrey and Vinyals, Oriol and Dean, Jeff},
  journal={arXiv preprint arXiv:1503.02531},
  year={2015}
}

@article{zagoruyko2016paying,
  title={Paying more attention to attention: Improving the performance of convolutional neural networks via attention transfer},
  author={Zagoruyko, Sergey and Komodakis, Nikos},
  journal={arXiv preprint arXiv:1612.03928},
  year={2016}
}

@inproceedings{sau2021deep,
  title={Deep Knowledge Distillation using Trainable Dense Attention.},
  author={Sau, Bharat Bhusan and Roy, Soumya and Namboodiri, Vinay P and Iyengar, Raghu Sesha},
  booktitle={BMVC},
  pages={72},
  year={2021}
}

@inproceedings{shin2022teaching,
  title={Teaching where to look: Attention similarity knowledge distillation for low resolution face recognition},
  author={Shin, Sungho and Lee, Joosoon and Lee, Junseok and Yu, Yeonguk and Lee, Kyoobin},
  booktitle={European Conference on Computer Vision},
  pages={631--647},
  year={2022},
  organization={Springer}
}

@inproceedings{jin2024align,
  title={Align-to-Distill: Trainable Attention Alignment for Knowledge Distillation in Neural Machine Translation},
  author={Jin, Heegon and Son, Seonil and Park, Jemin and Kim, Youngseok and Noh, Hyungjong and Lee, Yeonsoo},
  booktitle={Proceedings of the 2024 Joint International Conference on Computational Linguistics, Language Resources and Evaluation (LREC-COLING 2024)},
  pages={722--732},
  year={2024}
}

@inproceedings{lu2020twinbert,
  title={Twinbert: Distilling knowledge to twin-structured compressed bert models for large-scale retrieval},
  author={Lu, Wenhao and Jiao, Jian and Zhang, Ruofei},
  booktitle={Proceedings of the 29th ACM International Conference on Information \& Knowledge Management},
  pages={2645--2652},
  year={2020}
}

@inproceedings{lin2021batch,
  title={In-batch negatives for knowledge distillation with tightly-coupled teachers for dense retrieval},
  author={Lin, Sheng-Chieh and Yang, Jheng-Hong and Lin, Jimmy},
  booktitle={Proceedings of the 6th Workshop on Representation Learning for NLP (RepL4NLP-2021)},
  pages={163--173},
  year={2021}
}

@inproceedings{dong2023dual,
  title={Dual learning with dynamic knowledge distillation for partially relevant video retrieval},
  author={Dong, Jianfeng and Zhang, Minsong and Zhang, Zheng and Chen, Xianke and Liu, Daizong and Qu, Xiaoye and Wang, Xun and Liu, Baolong},
  booktitle={Proceedings of the IEEE/CVF International Conference on Computer Vision},
  pages={11302--11312},
  year={2023}
}

@article{ma2023using,
  title={Using multimodal contrastive knowledge distillation for video-text retrieval},
  author={Ma, Wentao and Chen, Qingchao and Zhou, Tongqing and Zhao, Shan and Cai, Zhiping},
  journal={IEEE Transactions on Circuits and Systems for Video Technology},
  volume={33},
  number={10},
  pages={5486--5497},
  year={2023},
  publisher={IEEE}
}

@article{rao2023dynamic,
  title={Dynamic contrastive distillation for image-text retrieval},
  author={Rao, Jun and Ding, Liang and Qi, Shuhan and Fang, Meng and Liu, Yang and Shen, Li and Tao, Dacheng},
  journal={IEEE Transactions on Multimedia},
  volume={25},
  pages={8383--8395},
  year={2023},
  publisher={IEEE}
}

@article{csizmadia2025distill,
  title={Distill CLIP (DCLIP): Enhancing Image-Text Retrieval via Cross-Modal Transformer Distillation},
  author={Csizmadia, Daniel and Codreanu, Andrei and Sim, Victor and Prabhu, Vighnesh and Lu, Michael and Zhu, Kevin and O'Brien, Sean and Sharma, Vasu},
  journal={arXiv preprint arXiv:2505.21549},
  year={2025}
}

@inproceedings{miech2021thinking,
  title={Thinking fast and slow: Efficient text-to-visual retrieval with transformers},
  author={Miech, Antoine and Alayrac, Jean-Baptiste and Laptev, Ivan and Sivic, Josef and Zisserman, Andrew},
  booktitle={Proceedings of the IEEE/CVF conference on computer vision and pattern recognition},
  pages={9826--9836},
  year={2021}
}

@article{izacard2020distilling,
  title={Distilling knowledge from reader to retriever for question answering},
  author={Izacard, Gautier and Grave, Edouard},
  journal={arXiv preprint arXiv:2012.04584},
  year={2020}
}

@inproceedings{zhou2024fine,
  title={Fine-grained distillation for long document retrieval},
  author={Zhou, Yucheng and Shen, Tao and Geng, Xiubo and Tao, Chongyang and Shen, Jianbing and Long, Guodong and Xu, Can and Jiang, Daxin},
  booktitle={Proceedings of the AAAI Conference on Artificial Intelligence},
  volume={38},
  number={17},
  pages={19732--19740},
  year={2024}
}

@article{li2024intermediate,
  title={Intermediate distillation: Data-efficient distillation from black-box llms for information retrieval},
  author={Li, Zizhong and Zhang, Haopeng and Zhang, Jiawei},
  journal={arXiv preprint arXiv:2406.12169},
  year={2024}
}

@article{beyer2024paligemma,
  title={Paligemma: A versatile 3b vlm for transfer},
  author={Beyer, Lucas and Steiner, Andreas and Pinto, Andr{\'e} Susano and Kolesnikov, Alexander and Wang, Xiao and Salz, Daniel and Neumann, Maxim and Alabdulmohsin, Ibrahim and Tschannen, Michael and Bugliarello, Emanuele and others},
  journal={arXiv preprint arXiv:2407.07726},
  year={2024}
}

@article{bai2025qwen2,
  title={Qwen2. 5-vl technical report},
  author={Bai, Shuai and Chen, Keqin and Liu, Xuejing and Wang, Jialin and Ge, Wenbin and Song, Sibo and Dang, Kai and Wang, Peng and Wang, Shijie and Tang, Jun and others},
  journal={arXiv preprint arXiv:2502.13923},
  year={2025}
}

@article{khosla2020supervised,
  title={Supervised contrastive learning},
  author={Khosla, Prannay and Teterwak, Piotr and Wang, Chen and Sarna, Aaron and Tian, Yonglong and Isola, Phillip and Maschinot, Aaron and Liu, Ce and Krishnan, Dilip},
  journal={Advances in neural information processing systems},
  volume={33},
  pages={18661--18673},
  year={2020}
}

@article{zhuang2024promptreps,
  title={Promptreps: Prompting large language models to generate dense and sparse representations for zero-shot document retrieval},
  author={Zhuang, Shengyao and Ma, Xueguang and Koopman, Bevan and Lin, Jimmy and Zuccon, Guido},
  journal={arXiv preprint arXiv:2404.18424},
  year={2024}
}

@article{yao2021filip,
  title={Filip: Fine-grained interactive language-image pre-training},
  author={Yao, Lewei and Huang, Runhui and Hou, Lu and Lu, Guansong and Niu, Minzhe and Xu, Hang and Liang, Xiaodan and Li, Zhenguo and Jiang, Xin and Xu, Chunjing},
  journal={arXiv preprint arXiv:2111.07783},
  year={2021}
}

@article{xie2025fg,
  title={FG-CLIP: Fine-Grained Visual and Textual Alignment},
  author={Xie, Chunyu and Wang, Bin and Kong, Fanjing and Li, Jincheng and Liang, Dawei and Zhang, Gengshen and Leng, Dawei and Yin, Yuhui},
  journal={arXiv preprint arXiv:2505.05071},
  year={2025}
}

@article{jing2024fineclip,
  title={Fineclip: Self-distilled region-based clip for better fine-grained understanding},
  author={Jing, Dong and He, Xiaolong and Luo, Yutian and Fei, Nanyi and Wei, Wei and Zhao, Huiwen and Lu, Zhiwu and others},
  journal={Advances in Neural Information Processing Systems},
  volume={37},
  pages={27896--27918},
  year={2024}
}

@inproceedings{radford2021learning,
  title={Learning transferable visual models from natural language supervision},
  author={Radford, Alec and Kim, Jong Wook and Hallacy, Chris and Ramesh, Aditya and Goh, Gabriel and Agarwal, Sandhini and Sastry, Girish and Askell, Amanda and Mishkin, Pamela and Clark, Jack and others},
  booktitle={International conference on machine learning},
  pages={8748--8763},
  year={2021},
  organization={PmLR}
}

@inproceedings{zhai2023sigmoid,
  title={Sigmoid loss for language image pre-training},
  author={Zhai, Xiaohua and Mustafa, Basil and Kolesnikov, Alexander and Beyer, Lucas},
  booktitle={Proceedings of the IEEE/CVF international conference on computer vision},
  pages={11975--11986},
  year={2023}
}

@article{darcet2023vision,
  title={Vision transformers need registers},
  author={Darcet, Timoth{\'e}e and Oquab, Maxime and Mairal, Julien and Bojanowski, Piotr},
  journal={arXiv preprint arXiv:2309.16588},
  year={2023}
}

@inproceedings{asokan2025finelip,
  title={FineLIP: Extending CLIP's Reach via Fine-Grained Alignment with Longer Text Inputs},
  author={Asokan, Mothilal and Wu, Kebin and Albreiki, Fatima},
  booktitle={Proceedings of the Computer Vision and Pattern Recognition Conference},
  pages={14495--14504},
  year={2025}
}

@article{bica2024improving,
  title={Improving fine-grained understanding in image-text pre-training},
  author={Bica, Ioana and Ili{\'c}, Anastasija and Bauer, Matthias and Erdogan, Goker and Bo{\v{s}}njak, Matko and Kaplanis, Christos and Gritsenko, Alexey A and Minderer, Matthias and Blundell, Charles and Pascanu, Razvan and others},
  journal={arXiv preprint arXiv:2401.09865},
  year={2024}
}

@inproceedings{mathew2021docvqa,
  title={Docvqa: A dataset for vqa on document images},
  author={Mathew, Minesh and Karatzas, Dimosthenis and Jawahar, CV},
  booktitle={Proceedings of the IEEE/CVF winter conference on applications of computer vision},
  pages={2200--2209},
  year={2021}
}

@inproceedings{mathew2022infographicvqa,
  title={Infographicvqa},
  author={Mathew, Minesh and Bagal, Viraj and Tito, Rub{\`e}n and Karatzas, Dimosthenis and Valveny, Ernest and Jawahar, CV},
  booktitle={Proceedings of the IEEE/CVF Winter Conference on Applications of Computer Vision},
  pages={1697--1706},
  year={2022}
}

@article{zhu2021tat,
  title={TAT-QA: A question answering benchmark on a hybrid of tabular and textual content in finance},
  author={Zhu, Fengbin and Lei, Wenqiang and Huang, Youcheng and Wang, Chao and Zhang, Shuo and Lv, Jiancheng and Feng, Fuli and Chua, Tat-Seng},
  journal={arXiv preprint arXiv:2105.07624},
  year={2021}
}

@article{li2024multimodal,
  title={Multimodal arxiv: A dataset for improving scientific comprehension of large vision-language models},
  author={Li, Lei and Wang, Yuqi and Xu, Runxin and Wang, Peiyi and Feng, Xiachong and Kong, Lingpeng and Liu, Qi},
  journal={arXiv preprint arXiv:2403.00231},
  year={2024}
}

@article{Abdin2024Phi3TR,
  title={Phi-3 Technical Report: A Highly Capable Language Model Locally on Your Phone},
  author={Marah Abdin and Sam Ade Jacobs and Ammar Ahmad Awan and Jyoti Aneja and Ahmed Awadallah and Hany Hassan Awadalla and Nguyen Bach and Amit Bahree and Arash Bakhtiari and Harkirat Singh Behl and Alon Benhaim and Misha Bilenko and Johan Bjorck and S{\'e}bastien Bubeck and Martin Cai and Caio C'esar Teodoro Mendes and Weizhu Chen and Vishrav Chaudhary and Parul Chopra and Allison Del Giorno and Gustavo de Rosa and Matthew Dixon and Ronen Eldan and Dan Iter and Abhishek Goswami and Suriya Gunasekar and Emman Haider and Junheng Hao and Russell J. Hewett and Jamie Huynh and Mojan Javaheripi and Xin Jin and Piero Kauffmann and Nikos Karampatziakis and Dongwoo Kim and Young Jin Kim and Mahoud Khademi and Lev Kurilenko and James R. Lee and Yin Tat Lee and Yuanzhi Li and Chen Liang and Weishung Liu and Eric Lin and Zeqi Lin and Piyush Madan and Arindam Mitra and Hardik Modi and Anh Nguyen and Brandon Norick and Barun Patra and Daniel Perez-Becker and Thomas Portet and Reid Pryzant and Heyang Qin and Marko Radmilac and Liliang Ren and Corby Rosset and Sambudha Roy and Olli Saarikivi and Amin Saied and Adil Salim and Michael Santacroce and Shital Shah and Ning Shang and Hiteshi Sharma and Xianmin Song and Olatunji Ruwase and Praneetha Vaddamanu and Xin Wang and Rachel Ward and Guanhua Wang and Philipp Andre Witte and Michael Wyatt and Can Xu and Jiahang Xu and Sonali Yadav and Fan Yang and Ziyi Yang and Donghan Yu and Cheng-Yuan Zhang and Cyril Zhang and Jianwen Zhang and Li Lyna Zhang and Yi Zhang and Yunan Zhang and Xiren Zhou and Yifan Yang},
  journal={ArXiv},
  year={2024},
  volume={abs/2404.14219}
}

@misc{xu2025llamanemoretrievercolembedtopperforming,
      title={Llama Nemoretriever Colembed: Top-Performing Text-Image Retrieval Model}, 
      author={Mengyao Xu and Gabriel Moreira and Ronay Ak and Radek Osmulski and Yauhen Babakhin and Zhiding Yu and Benedikt Schifferer and Even Oldridge},
      year={2025},
      eprint={2507.05513},
      archivePrefix={arXiv},
      primaryClass={cs.CV},
      url={https://arxiv.org/abs/2507.05513}, 
}

\end{document}